\documentclass[10pt,journal,compsoc]{IEEEtran}
\usepackage{graphicx}
\usepackage{amsmath}
\usepackage{multirow}
\usepackage{amssymb}
\usepackage{amsmath}
\usepackage{multirow}
\usepackage{colortbl}
\usepackage[dvipsnames]{xcolor}
\usepackage{hyperref}
\hypersetup{
    colorlinks=true,
    linkcolor=blue,
    filecolor=magenta,      
    urlcolor=cyan,
}

\ifCLASSOPTIONcompsoc
  \usepackage[nocompress]{cite}
\else
  \usepackage{cite}
\fi

\ifCLASSINFOpdf
\else
\fi

\begin{document}

\title{NeuroNet: A Novel Hybrid Self-Supervised Learning Framework  for Sleep Stage Classification Using Single-Channel EEG}

\author{
    Cheol-Hui Lee,
    Hakseung~Kim,
    Hyun-jee~Han,
    Min-Kyung Jung,
    Byung C. Yoon and Dong-Joo Kim

\IEEEcompsocitemizethanks{
    \IEEEcompsocthanksitem Cheol-Hui Lee and Min-Kyung Jung are with the Department of Brain and Cognitive Engineering, Korea University, Seoul, South Korea; with Interdisciplinary Program in Precision Public Health, Korea University, Seoul, South Korea
    \IEEEcompsocthanksitem Hakseung Kim is with the Department of Brain and Cognitive Engineering, Korea University, Seoul, South Korea
    \IEEEcompsocthanksitem Hyun-jee Han is with the Department of Pharmacology, University of Cambridge, Cambridge, UK
    \IEEEcompsocthanksitem Byung C. Yoon is with the Department of Radiology, Stanford University School of Medicine, VA Palo Alto Health Care System, Palo Alto, CA, USA
    \IEEEcompsocthanksitem Dong-Joo Kim is with the Department of Brain and Cognitive Engineering, Korea University, Seoul, South Korea; with the Department of Neurology, Korea University College of Medicine, Seoul, South Korea; with Interdisciplinary Program in Precision Public Health, Korea University, Seoul, South Korea
    
    E-mail: dongjookim@korea.ac.kr}
    \thanks{preprint}
}

\markboth{Journal of \LaTeX\ Class Files}%
{Shell \MakeLowercase{\textit{et al.}}: Bare Advanced Demo of IEEEtran.cls for IEEE Computer Society Journals}

\IEEEtitleabstractindextext{%
\begin{abstract}
The classification of sleep stages is a pivotal aspect of diagnosing sleep disorders and evaluating sleep quality. However, the conventional manual scoring process, conducted by clinicians, is time-consuming and prone to human bias. Recent advancements in deep learning have substantially propelled the automation of sleep stage classification. Nevertheless, challenges persist, including the need for large datasets with labels and the inherent biases in human-generated annotations. This paper introduces NeuroNet, a self-supervised learning (SSL) framework designed to effectively harness unlabeled single-channel sleep electroencephalogram (EEG) signals by integrating contrastive learning tasks and masked prediction tasks. NeuroNet demonstrates superior performance over existing SSL methodologies through extensive experimentation conducted across three polysomnography (PSG) datasets. Additionally, this study proposes a Mamba-based temporal context module to capture the relationships among diverse EEG epochs. Combining NeuroNet with the Mamba-based temporal context module has demonstrated the capability to achieve, or even surpass, the performance of the latest supervised learning methodologies, even with a limited amount of labeled data. This study is expected to establish a new benchmark in sleep stage classification, promising to guide future research and applications in the field of sleep analysis. The source code is available at \url{https://github.com/dlcjfgmlnasa/NeuroNet}

\end{abstract}

\begin{IEEEkeywords}
self-supervised learning, electroencephalogram (EEG), polysomnography, automatic sleep staging
\end{IEEEkeywords}}

\maketitle

\IEEEdisplaynontitleabstractindextext
\IEEEpeerreviewmaketitle

\ifCLASSOPTIONcompsoc
\IEEEraisesectionheading{\section{Introduction}\label{sec:introduction}}
\else
\section{Introduction}
\label{sec:introduction}
\fi

\IEEEPARstart{S}{leep} constitutes a fundamental determinant of human health and lifespan, serving as a cornerstone in alleviating both mental and physical stress encountered during routine activities, while also contributing to the maintenance of physiological homeostasis \cite{RN1}. However, many individuals suffer from sleep disorders \cite{RN42}, and polysomnography (PSG) is commonly employed to assess sleep quality as a part of their treatment. PSG entails a measurement of various physiological signals during sleep, including electroencephalogram (EEG), electromyography, and electrocardiogram \cite{RN2, RN3} that require a laborious process of manually analyzing the data and classifying sleep stages by sleep experts.

\begin{figure}[!htbp]
    \centerline{\includegraphics[width=\columnwidth]{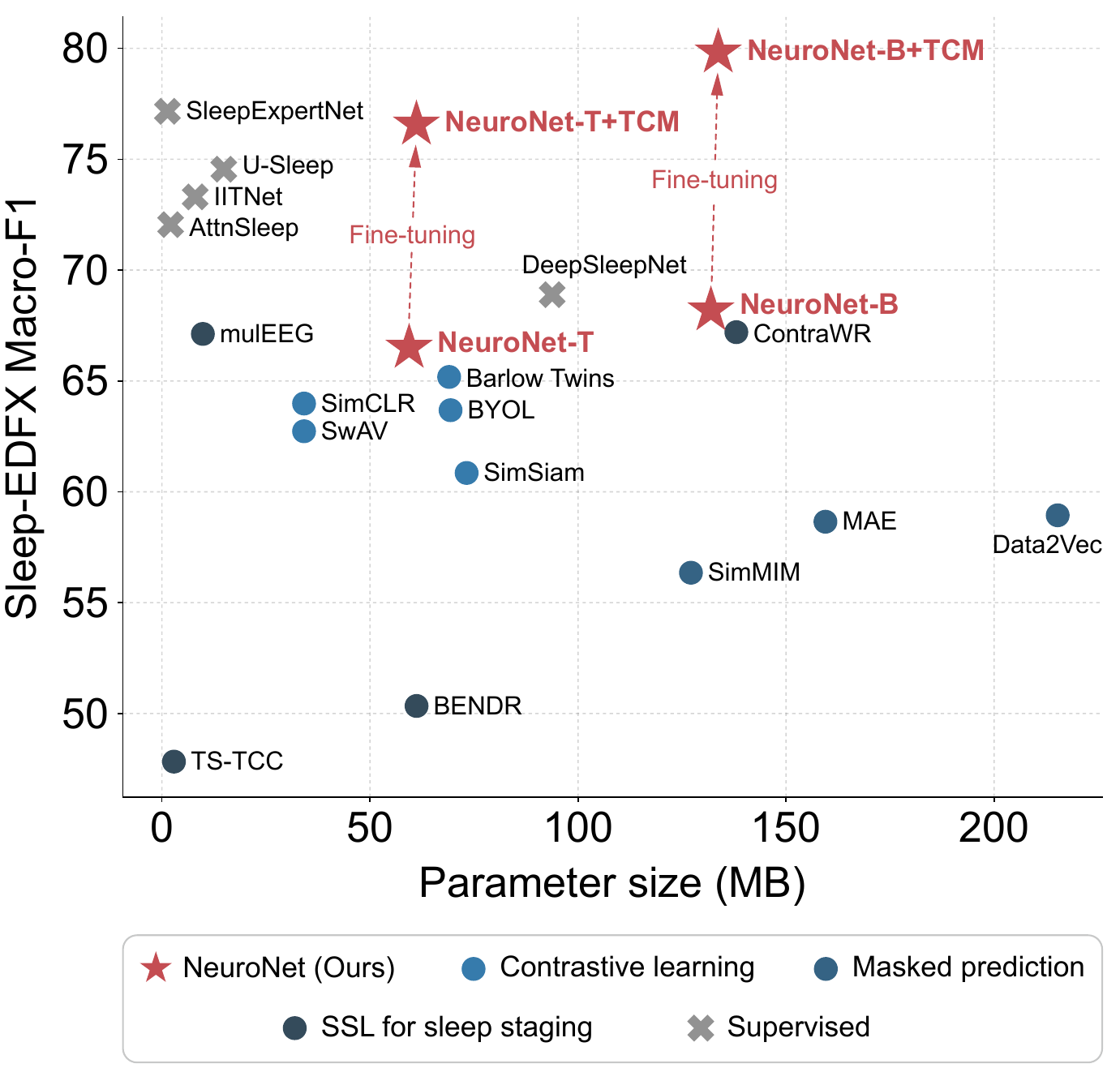}}
    \caption{Performance of Sleep-EDFX across various self-supervised learning and supervised learning.}
    \label{fig:figure_1} 
\end{figure}

Given these contexts, research into automatic sleep stage classification is advancing, with studies exclusively utilizing single channel EEG gaining particular attention due to their user convenience. The majority of these studies are based on supervised learning methodologies and have demonstrated acceptable performance enhancements through the utilization of the latest deep learning algorithms. Nonetheless, the implementation of such methodologies in real-world settings can pose several challenges. Firstly, the necessity for an extensive amount of labeled data may render its acquisition impractical. Secondly, the reliability of labels is compromised by low inter-rater agreement rates observed in sleep stage evaluation \cite{RN43}. Lastly, a model trained on data annotated by a single evaluator is prone to bias towards the opinions and interpretations of that evaluator, thereby posing a substantial risk to its generalizability.

Self-supervised learning (SSL) \cite{RN44} has emerged as a promising methodology for extracting meaningful representations from unlabeled data. SSL comprises two principal paradigms: masked prediction tasks, which aim to learn the intrinsic feature information of the data, and contrastive learning, which trains models to distinguish between similar and dissimilar pairs of data points. Serving as a pre-training method, SSL trains models using pseudo-labels derived from the inherent features or similarities present in the data. Models pre-trained via SSL can subsequently be fine-tuned even using a small amount of labeled data. This approach not only enables the effective utilization of large volumes of unlabeled data but also mitigates the issues of reduced generalization resulting from inaccurate or biased labels.

The attributes of SSL render it suitable for applications in sleep EEG, where data acquisition is straightforward yet labeling proves challenging. Despite its potential, SSL research in sleep EEG remains relatively limited. A review of existing studies reveals that the majority of studies rely on contrastive learning tasks \cite{RN38, RN28, RN30}. These studies have successfully learned representations of EEG, yet their effectiveness significantly depends on the backbone network's performance and on EEG data augmentation techniques \cite{RN28, RN30}. Although masked prediction tasks are not yet prevalent in sleep EEG research, their straightforward architecture, along with the capability to learn rich representations, has garnered popularity in other disciplines, such as computer vision. Nonetheless, they are perceived as less effective at learning discriminative representations, potentially leading to diminished performance in the downstream tasks \cite{RN41, RN40}.

This study proposes a model, NeuroNet, a pioneering SSL framework that effectively integrates the capacity to discern inherent information within the dataset via masked prediction tasks with the discriminative representation capabilities afforded by contrastive learning tasks. Additionally, a deep learning methodology known as Mamba \cite{RN45} is employed in this study, based on the selective state space model as an alternative to Long Short-Term Memory (LSTM) or multi-head attention, which are the conventional components of the temporal context module (TCM) utilized for integrating information across various time-zone EEG epochs. 

To the best of our knowledge, the proposed methodology represents an innovative approach yet to be explored within the domain of EEG artificial intelligence research. Our findings underscore the superior performance of NeuroNet over existing SSL methodologies. Furthermore, this study unveils that upon fine-tuning, the model leveraging pretrained NeuroNet in conjunction with Mamba \cite{RN45} outperforms the latest supervised learning technique trained on extensive labeled datasets even when utilizing only a limited amount of labeled datasets.

\section{Related Works}
\subsection{Sleep Stage Classification}

In recent years, research on deep learning-based sleep stage classification has been prolific. Notably, H. Phan et al. have proposed several models for sleep stage classification, as evidenced by several studies \cite{RN5, RN6, RN7}, which involve the conversion of EEG, electrooculography, and electromyography signals into time-frequency images for utilization as inputs to the models. Specifically, Multi-task CNN \cite{RN5} employs a convolutional neural network (CNN) architecture incorporating two convolutional layers and two max-pooling layers. SeqSleepNet \cite{RN6} adopts a Seq2Seq architecture, employing bidirectional long short-term memory (bi-LSTM). Similarly, XSleepNet \cite{RN7} adheres to the Seq2Seq model structure similar to SeqSleepNet \cite{RN6}; however, it distinctively integrates raw EEG signals as an additional input, setting it apart from Multi-task CNN \cite{RN5} and SeqSleepNet \cite{RN6}. XSleepNet \cite{RN7} implements a strategic approach that dynamically adjusts the learning rate, increasing it during periods of effective generalization and decreasing it to prevent overfitting when necessary.

For the sake of user convenience, there are studies that opt to utilize solely single-channel EEG. Both DeepSleepNet \cite{RN9} and IITNet \cite{RN10} extract representation vectors from EEG signals using CNNs and learn temporal context information through bi-LSTM. While DeepSleepNet \cite{RN9} focuses on temporal context between inter-epochs, IITNet \cite{RN10} comprehensively considers temporal context within and between epochs. Recently, multi-head attention has emerged as a primary alternative to Recurrent Neural Network (RNN)-based models for capturing temporal dependencies swiftly and efficiently. Notable amongst those is AttnSleep \cite{RN11} which is capable of extracting low-frequency and high-frequency features of EEG through multi-resolution CNNs, while adaptive feature recalibration enhances the quality of extracted features by modeling interdependencies between features. Subsequently, multi-head attention is employed to capture temporal dependencies among features. SleepExpertNet \cite{RN12} extracts representation vectors of EEG via spectral-temporal CNN after signal extraction from disparate frequency bands. This is followed by the implementation of a model combining multi-head attention with bi-LSTM is utilized to learn long- and short-term temporal context information.

\subsection{Self-Supervised Learning}
SSL, a framework designed to extract highly semantic patterns directly from data without relying on labels, operates through a two-stage pipeline. In the initial stage, it learns generated pseudo-labels via arbitrarily defined tasks. Subsequently, in the second stage, supervised learning is conducted using data with a limited number of labels. For this reason, SSL can be characterized as an intermediate approach between unsupervised and supervised learning methodologies. Currently, SSL research primarily revolves around two principal paradigms: contrastive learning tasks and masked prediction tasks.
    
\subsubsection{Contrastive Learning Task}
The objective of the contrastive learning task is to elucidate the interrelationship among multiple samples, employing various methodologies (e.g., negative samples, self-distillation, clustering, and feature decorrelation, etc.). Negative sampling aims to minimize the distance between positive pairs in the latent space while increasing the distance between negative pairs \cite{RN13, RN14}. 
To achieve this, a large number of contrastive pairs are required, and methods like MOCO \cite{RN13} and SimCLR \cite{RN14} utilize memory banks and large batch sizes, respectively. Self-distillation trains online networks to predict the output values of target networks, thereby obviating reliance on negative samples \cite{RN15, RN16}. BYOL \cite{RN15} and SimSiam \cite{RN16}, both based on self-distillation, present analogous structures. Nonetheless, BYOL \cite{RN15} involves updating the encoder used in the target network with a momentum encoder during training. In contrast, SimSiam \cite{RN16} does not use a momentum encoder but employs stop-gradient techniques instead. SwAV \cite{RN17}, a prominent example of clustering, trains representation vectors generated from identical samples to forecast the same prototype class. Feature decorrelation aims to learn decorrelated features \cite{RN18, RN19}. Barlow Twins \cite{RN18} computes the cross-correlation matrix between outputs of identical networks and trains it to align as closely as feasible with an identity matrix.
    
\subsubsection{Masked Prediction Task}
The masked prediction task is a method where parts of an data sample are masked and then restored to learn representations. Examining SSL methodologies based on the masked prediction task, MAE \cite{RN20} employs a ViT-based asymmetric autoencoder structure. Through the implementation of masking, MAE \cite{RN20} selectively transforms a segment of the input image into a latent vector via the encoder and then trains the decoder to reconstruct the original image. To minimize spatial redundancy, a substantial proportion of masking (75\% or higher) is applied, leading to notable enhancements in generalization performance. BEiT \cite{RN21} initiates pre-training with a vector quantized-variational autoencoder, subsequently employing it as a tokenizer. It further segments the image into patches and leverages a subset of the masked patches as input for the ViT-based encoder, training it to predict the originally masked content based on the tokenizer. SimMIM \cite{RN22} adopts a comparable structure to BEiT \cite{RN21}, yet it opts for pixel regression tasks over intricate methodologies like tokenization or clustering. Data2Vec \cite{RN23} utilizes a teacher network trained on original data and a student network trained on masked data to create representation vectors. The student networks are trained to predict the representation vectors of the target network.
    
\subsubsection{Hybrid Approach: Contrastive Learning Task + Masked Prediction Task}
Recent research has initiated discussions on the combination of contrastive learning tasks and masked prediction tasks. CMAE \cite{RN40} employs a masked autoencoder structure in the online branch and conducts the task of reconstructing the original image. The target branch employs a momentum-updated encoder and receives the entire image to conduct contrastive learning in tandem with the online branch. This model incorporates pixel shifting as a form of data augmentation. CAN \cite{RN41} also combines contrastive learning and masked autoencoder structures similar to CMAE \cite{RN40}. Noteworthy is its incorporation of noise prediction, which distinguishes it and enhances its capacity to acquire more refined representations.

\subsection{Self-Supervised Learning for Sleep Staging}
BENDR \cite{RN25} integrates a CNN-based module for EEG signal learning alongside a Transformer-based module for learning temporal context between signals. In BENDR \cite{RN25}, output vectors extracted through the CNN and Transformer modules are defined as positive pairs if they are from the same time point, and as negative pairs if they originate from distinct time points. Subsequently, training of these pairs is conducted utilizing InfoNCE loss. ContraWR \cite{RN26} opts for triplet loss over the conventional InfoNCE loss for contrastive learning. This strategy enables minimization of the distance between positive pairs as well as simultaneous increases in the gap between negative pairs. Negative samples for each sample are substituted with the average value, termed the world representation vector. CoSleep \cite{RN27} adopts a multi-view SSL approach to concurrently learn signals and spectrogram images. It incorporates a module comprising a queue and a momentum encoder to secure a multitude of negative pairs, with the goal of augmenting representation performance. TS-TCC \cite{RN28} focuses on temporal representation learning via a temporal contrasting module using weak and strong augmentations applied EEG signals. It maximizes the similarity between contexts originating from the same sample while minimizing the similarity between contexts of different samples. MAEEG \cite{RN29} engages in representation learning for 6-channel sleep EEG using a masked autoencoder. Fine-tuned MAEEG presents enhanced performance in sleep stage classification even with limited labels provided. mulEEG \cite{RN30} adopts the same augmentation approach as TS-TCC and has a multi-view SSL structure. It employs EEG signals and spectrograms transformed from EEG signals as input data, implementing diverse loss functions to effectively learn complementary information from multiple views.

\begin{figure*}[!t]
    \centerline{\includegraphics[width=0.85\textwidth]{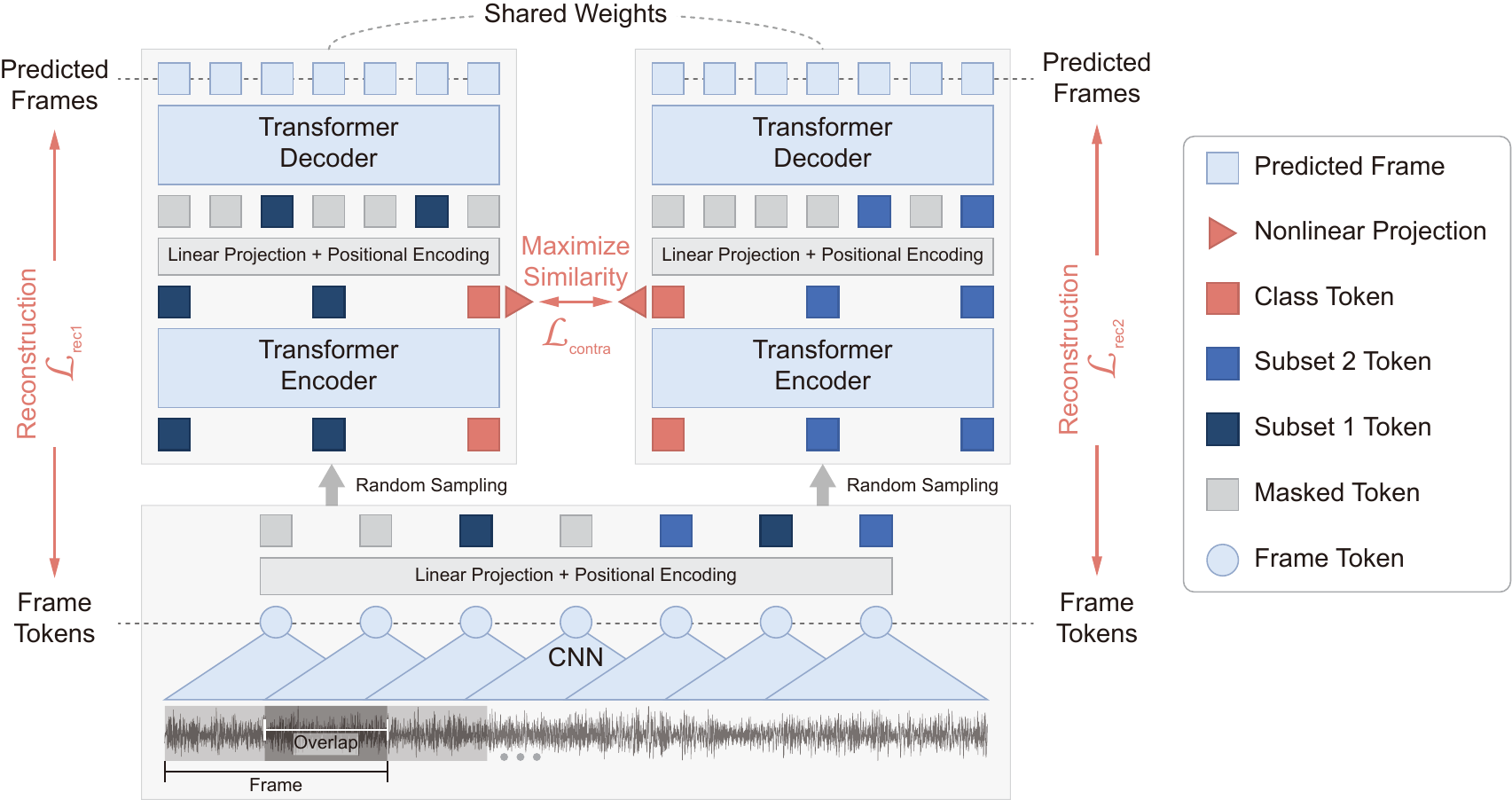}}
    \caption{Overview of the NeuroNet framework architecture.}    
    \label{fig:figure_2}
\end{figure*}

\section{Methodology}
In this section, we introduce NeuroNet and the Mamba-based TCM. Figure \ref{fig:figure_2} illustrates the detailed framework of NeuroNet. NeuroNet is an SSL framework designed to learn EEG signals, consisting of a total of five training stages. TCM is utilized to effectively capture time-series features or relational information between multiple EEG epochs, akin to a sleep expert. The TCM learns by decoding the intricate temporal patterns inherent within EEG signal, taking the output vector from the pretrained NeuroNet encoder as its input.

\subsection{NeuroNet: Contrastive Masked Autoencoder for EEG}
\subsubsection{Data Preprocessing}
Prior to training the model, the EEG signals were bandpass filtered between 1 and 50 Hz and then resampled at a frequency of 100 Hz. Following this preprocessing, a sliding window methodology was applied to each EEG epoch signal. The sliding window is a technique that involves moving a fixed-size frame at regular intervals across the data, facilitating the analysis which is effectively utilized in identifying specific patterns or events embedded within. The signals extracted through the sliding window served as the input data for the frame network.
    
\subsubsection{Frame Network}
The frame network is designed based on a multiscale 1d ResNet \cite{RN39}, specifically tailored for time series classification tasks. This model comprises an initial shared convolution and three parallel feature extractors. The shared convolution includes a 1d convolution layer, batch normalization, and a max pooling layer. Each feature extractor consists of three convolution blocks, with each block including a convolution layer, batch normalization, and an Exponential Linear Unit (ELU) activation function. Following the final block, residual connections and average pooling are applied. Notably, each feature extractor differs by employing 1d convolution layers with varying kernel sizes, which are 3, 5, and 7 respectively. The feature vectors extracted from distinct feature extractors are concatenated and subsequently processed through two fully connected layers to produce the final vector $ {\left\{{z_i}^{m}\right\}}_{m=1}^{M} $, where $M $ represents the total number of frames, and $i $ is the frame index.
    
\subsubsection{Masked Prediction Task}
The structure of the MAE has been utilized for the masked prediction task. Figure \ref{fig:figure_2} illustrates that the input data ${\left\{{z_i}^{m}\right\}}_{m=1}^{M} $ undergoes two separate masked prediction tasks independently. This architecture encompasses an encoder responsible for mapping the input to latent representations and a decoder that reconstructs the original signal from the latent representations. The encoder operates exclusively on the observed portion of the signal, without any mask tokens, while the decoder reconstructs the entire signal from the latent representation and mask tokens.

\textbf{(Masking)} Given the vector ${\left\{{z_i}^{m}\right\}}_{m=1}^{M} $ extracted through the frame network, a small subset of frames is randomly sampled for training, while the remainder are masked. This subset is denoted as ${\left\{{z_i}^{m}\right\}}_{m=1}^{\widetilde M} $, where $\widetilde M $ represents the frame numbers of the sampled subset. Generally, as the masking ratio increases, the amount of information available to the model decreases, which raises the difficulty of the reconstruction task. This enables models to understand the underlying patterns of the data and allows for a more generalized representation of the input data.

\textbf{(Encoder)} The encoder within NeuroNet is charged with the pivotal task of encoding the vector extracted via the frame layer into tokens and subsequently mapping them to the latent space. In NeuroNet, a standard Transformer stack serves as the encoder. The multi-head attention mechanism of the Transformer excels in efficiency capturing temporal information and correlations among frames. The encoder receives as input a vector obtained by concatenating the class token and ${\left\{{z_i}^{m}\right\}}_{m=1}^{\widetilde M} $, and then applying `linear projection + positional encoding'. Through the Transformer stack, it produces a new vector ${\left\{{h_i}^{m}\right\}}_{m=1}^{\widetilde M+1} $. The class token assumes a crucial role in extracting semantic information from the data, thereby proving instrumental in contrastive learning tasks and predictions.

\textbf{(Decoder)} The decoder receives the vectors ${\left\{{h_i}^{m}\right\}}_{m=1}^{\widetilde M} $ extracted by the encoder, excluding the class token, and the masked vectors as inputs. Similar to the encoder, it applies linear projection and positional encoding to the input vectors. The masked vectors serve to represent vectors that were excluded during the masking phase, encapsulating the information omitted from the input data. In alignment with the asymmetric nature of the MAE, a tiny standard Transformer-based decoder is utilized. However, the decoder is not engaged after the SSL phase.

\textbf{(Reconstruction Target)} \textbf{\space }To predict the values of the masked vectors, the output vector generated by the decoder is passed through a fully connected layer to be reconstructed to the same size as the vector ${\left\{{z_i}^{m}\right\}}_{m=1}^{M} $\noindent  extracted through the frame layer. Through this process, NeuroNet is able to derive the reconstruction vector ${\left\{{r_i}^{m}\right\}}_{m=1}^{M}\; $\noindent , and the loss is calculated using the mean square error, but only for the masked vectors. The formula for the loss function is as follows: 
\begin{equation}
\label{equation_1}
L_{rec}=\frac1{N\left(M-\widetilde M\right)}{\sum_{i=1}^{N}{{\sum_{m=1}^{M-\widetilde M}{{{(z_i}^{m}-{r_i}^{m})}^{2}}}}}
\end{equation}
Here, $M$ and $N$ represent the total number of data points and the batch size, respectively. As shown in Figure \ref{fig:figure_2}, since NeuroNet performs two independent masked prediction tasks using a single encoder and decoder, two separate losses (e.g., $L_{rec1} $, $L_{rec2} $) are ultimately derived.
    
\subsubsection{Contrastive Learning Task}
The objective of the contrastive learning task is to learn valuable representations by maximizing the similarity amongst identical instances while concurrently minimizing the similarity between disparate instances. In NeuroNet, the NT-Xent loss \cite{RN50} is leveraged, which trains the model to converge positive pairs closer together and simultaneously drive negative pairs further apart within each mini-batch.

In Figure \ref{fig:figure_2}, random sampling results in two distinct subsets that are fed into the encoder. From this process, vectors $h_i $ and $h_j $ representing different views are derived through the class token. After mapping $h_i $ and $h_j $ through the projection layer to a new latent space, the normalized $c_i $, $c_j $ are obtained. If the batch size is $N $, then $c_i $ and $c_j $ are each generated in sets of $N $. Consequently, each sample can generate 1 positive pair and $2\left(N-1\right) $ negative pairs. Thus, by iterating from $k=1 $ to $k=2N $, and avoiding references to the same sample for both $k $ and $i $, the formula is as follows:
\begin{equation}
\label{equation_2}
L_{contra}=\frac1{2N}{\sum_{k=1}^{N}{\left[l\left(2k-1,\;2k\right)+l\left(2k,2k-1\right)\right]}}
\end{equation}

\begin{equation}
\label{equation_3}
l(i,j)={{-log}{\frac{exp(\text{sim}\left(c_i,\;c_j\right)/\tau)}{{\sum_{k=1}^{2N}{1_{\lbrack k\neq i\rbrack}exp(\text{sim}\left(c_i,\;c_k\right)/\tau)}}}}}
\end{equation}
Here, $\tau>0\; $ represents the temperature, a hyperparameter, with a default setting of $\tau=0.5 $. The term $sim $ denotes cosine similarity.
    
\subsubsection{The Combined Loss Function}
NeuroNet aims to learn more superior representations by combining the masked prediction task, which learns the semantic information of EEG signals, with the contrastive learning task, which learns the relationships between EEG signals. The formula is as follows:
\begin{equation}
\label{equation_4}
L_{total}=\;\frac12\left(L_{rec1}+L_{rec2}\right)+\alpha L_{contra}
\end{equation}
Here, $\alpha $ is the hyperparameter that balances the two types of loss mentioned above.

\subsection{Mamba-based Temporal Context Modules}
In the landscape of sequence modelling, previous models such as LSTM and multi-head attention have been situated within a trade-off between effectiveness and efficiency. Recently, Mamba \cite{RN45}, a structured state space sequence model featuring a selective mechanism and scan module, has emerged as a powerful tool in long sequence modeling. The selective mechanism employs input-dependent parameters, unlike space state models that utilize constant transition parameters. This adaptive strategy enhances overall generalization and performance by selectively focusing on relevant information while disregarding noisy or extraneous data. Complementing this, the scan module is deployed across each window of the input sequence, adeptly capturing intricate patterns and dependencies spanning multiple time steps. Additionally, a hardware-aware algorithm facilitates linear expansion of sequence length, thereby optimizing and improving computational efficiency and resource utilization. 

According to the American Academy of Sleep Medicine (AASM) \cite{RN51}, when classifying sleep stages, not only the local features occurring in a single-epoch EEG (e.g., K-complex, sleep spindle, etc.) are considered, but also the relationships between adjacent EEG epochs are comprehensively considered to determine the stage of sleep. In this study, a Mamba-based TCM was developed to efficiently capture the temporal characteristics and correlations among multiple EEG epochs. Concretely, sleep stages are classified by a Mamba-based TCM that receives class tokens derived through an encoder from multiple EEG epochs. This process can be formula expressed as follows:
\begin{equation}
\label{equation_5}
Sleep\;Stage=Mamba({\{{Cls\;Token}_n\}}_{n=1}^{N})
\end{equation}

\begin{equation}
\label{equation_6}
Cls\;Token=Encoder(EEG\;Epoch)
\end{equation}
Here, $N$ represents the number of EEG epochs input simultaneously, which is $N$ is 20.

\section{Experiments}
\subsection{Dataset Description} In this study, three PSG datasets were utilized.

        \begin{table*}
        \caption{Experimental settings and dataset statistics.}
        \label{table:table_1}
        \centering
        \renewcommand{\arraystretch}{1.145}
        \resizebox{\textwidth}{!}{
        \begin{tabular}{c|ccccc|cccccc} 
        \hline
        \multirow{2}{*}{\textbf{\textit{Dataset}}} & \multirow{2}{*}{\begin{tabular}[c]{@{}c@{}}\textbf{\textit{No. of }}\\\textbf{\textit{subjects}}\end{tabular}} & \multirow{2}{*}{\begin{tabular}[c]{@{}c@{}}\textbf{\textit{EEG}}\\\textbf{\textit{Channel}}\end{tabular}} & \multirow{2}{*}{\begin{tabular}[c]{@{}c@{}}\textbf{\textit{Evaluation}}\\\textbf{\textit{Scheme}}\end{tabular}} & \multirow{2}{*}{\begin{tabular}[c]{@{}c@{}}\textbf{\textit{Held-out}}\\\textbf{\textit{Validation Set}}\end{tabular}} & \multirow{2}{*}{\begin{tabular}[c]{@{}c@{}}\textbf{\textit{Sampling}} \\\textbf{\textit{Rate}}\end{tabular}} & \multicolumn{6}{c}{\textbf{\textit{Class Distribution}}}                                                                                                                                                                                                                                                                                                                                                                           \\ 
        \cline{7-12}
                                                   &                                                                                                                &                                                                                                           &                                                                                                                 &                                                                                                                       &                                                                                                              & \textbf{\textit{Wake}}                                                      & \textbf{\textit{N1}}                                                        & \textbf{\textit{N2}}                                                         & \textbf{\textit{N3}}                                                        & \textbf{\textit{REM}}                                                       & \textbf{\textit{\# Total}}  \\ 
        \hline
        \textit{Sleep-EDFX\textbf{}}               & 153\textbf{\textit{}}                                                                                          & Fpz-Cz\textbf{\textit{}}                                                                                  & 5-fold CV\textbf{\textit{}}                                                                                     & 15
          subjects                                                                                                         & 100 Hz\textbf{\textit{}}                                                                                        & \begin{tabular}[c]{@{}c@{}}66822 \\(34.78\%)\textbf{\textit{}}\end{tabular} & \begin{tabular}[c]{@{}c@{}}21522 \\(11.20\%)\textbf{\textit{}}\end{tabular} & \begin{tabular}[c]{@{}c@{}}69132 \\(35.99\%)\textbf{\textit{}}\end{tabular}  & \begin{tabular}[c]{@{}c@{}}8793 \\(4.58\%)\textbf{\textit{}}\end{tabular}   & \begin{tabular}[c]{@{}c@{}}25835 \\(13.45\%)\textbf{\textit{}}\end{tabular} & 192104\textbf{\textit{}}    \\
        \textit{SHHS\textbf{}}                     & 329\textbf{\textit{}}                                                                                          & C4-A1\textbf{\textit{}}                                                                                   & 5-fold CV\textbf{\textit{}}                                                                                     & 40
          subjects                                                                                                         & 125 Hz\textbf{\textit{}}                                                                                        & \begin{tabular}[c]{@{}c@{}}59129 \\(17.51\%)\textbf{\textit{}}\end{tabular} & \begin{tabular}[c]{@{}c@{}}10304 \\(3.05\%)\textbf{\textit{}}\end{tabular}  & \begin{tabular}[c]{@{}c@{}}142125 \\(42.09\%)\textbf{\textit{}}\end{tabular} & \begin{tabular}[c]{@{}c@{}}60153 \\(17.81\%)\textbf{\textit{}}\end{tabular} & \begin{tabular}[c]{@{}c@{}}65953 \\(19.53\%)\textbf{\textit{}}\end{tabular} & 337664\textbf{\textit{}}    \\
        \textit{ISRUC-Sleep\textbf{}}              & 100\textbf{\textit{}}                                                                                          & C4-A1\textbf{\textit{}}                                                                                   & 10-fold CV\textbf{\textit{}}                                                                                    & 10
          subjects                                                                                                         & 200 Hz\textbf{\textit{}}                                                                                        & \begin{tabular}[c]{@{}c@{}}22142 \\(24.55\%)\textbf{\textit{}}\end{tabular} & \begin{tabular}[c]{@{}c@{}}9140 \\(10.13\%)\textbf{\textit{}}\end{tabular}  & \begin{tabular}[c]{@{}c@{}}30499 \\(33.82\%)\textbf{\textit{}}\end{tabular}  & \begin{tabular}[c]{@{}c@{}}16115 \\(17.87\%)\textbf{\textit{}}\end{tabular} & \begin{tabular}[c]{@{}c@{}}12291 \\(13.63\%)\textbf{\textit{}}\end{tabular} & 90187\textbf{\textit{}}     \\
        \hline
        \multicolumn{4}{l}{*\footnotesize{ CV: Cross Validation}}
        \end{tabular}}
        \end{table*}
    
\subsubsection{Sleep-EDF Expanded Dataset}

This study utilized the Sleep-EDF expanded dataset (Sleep-EDFX) \cite{RN31}, which is divided into two subsets: SC and ST. The SC subset contains PSG recordings from 153 healthy individuals aged 25 to 101, designed to explore the effects of age on sleep patterns. On the other hand, the ST subset comprises PSG recordings from 44 individuals aged 18 to 79, specifically curated to examine the impact of temazepam on sleep. The PSG configuration includes two bipolar EEG channels (Fpz-Cz and Pz-Oz), a horizontal EOG channel, and a submental chin EMG channel. Sleep stages were classified every 30 seconds by sleep experts into one of eight categories: \{'W', 'N1', 'N2', 'N3', 'N4', 'REM', 'M', and '?'\}. In this research, only the SC subset was utilized, selecting the Fpz-Cz EEG signal with a sampling rate of 100 Hz. Additionally, in alignment with the AASM standard, out of the 8 classes, N3 and N4 were merged into N3, and the classes 'M' and '?' were omitted.
    
\subsubsection{Sleep Heart Health Study}
The Sleep Heart Health Study (SHHS) \cite{RN32, RN33} is a comprehensive multi-center cohort study designed to investigate various cardiovascular and other outcomes associated with sleep-disordered breathing. It consists of two subsets: SHHS1, SHHS2. Each subset within the PSG includes two bipolar EEG channels (C4-A1, C3-A2), one EKG channel, two EOG channels, as well as two lower limb EMG channels, snoring detection, pulse oximeters, and a body position sensor. Sleep experts labeled the recordings every 30 seconds into one of eight categories: \{'W', 'N1', 'N2', 'N3', 'N4', 'REM', 'Movement', and 'Unknown'\}. In this study, 329 participants from the SHHS1 subset, considered to have regular sleep patterns (Apnea Hypopnea or AHI index less than 5) \cite{RN34}, were selected. The C4-A1 EEG signal with a sampling rate of 125 Hz was chosen. Following the AASM standard, the classes N3 and N4 were merged into N3, and 'Movement' and 'Unknown' were omitted.
    
\subsubsection{ISRUC Sleep Dataset}
The ISRUC-Sleep dataset \cite{RN35} consists of 3 subsets and was collected to study both healthy subjects and those taking sleep medication. The first subset comprises data from 100 participants, each with only one PSG recorded. The second subset includes data from 8 participants, each with two PSG sessions. The third subset comprises data from 10 healthy participants, each with only one PSG recorded. This subset proves particularly useful for conducting comparative analyses between healthy participants and individuals afflicted with sleep disorders. In this dataset, two sleep experts labeled each 30-second interval with one of the five classes \{`W', `N1', `N2', `N3', `REM'\}. This study utilized the C4-A1 EEG signal from the first subset, sampled at a rate of 200 Hz, and employed the labeling provided by the first sleep expert.

\subsection{Other State-of-the-Art Methodologies} Among various studies focusing on sleep stage classification and SSL methods, several with methodologies of significant influence (i.e., a high number of citations) and available published source code were selected and implemented. Such selection underscores the preference for methodologies with high reproducibility, which have been validated by numerous researchers in the field.  The selected recent methodologies can be categorized into contrastive learning-based SSL, masked prediction task-based SSL, SSL for sleep staging, and supervised learning for sleep staging. 

Firstly, for the contrastive learning-based SSL methodology, SimCLR \cite{RN14}, BYOL \cite{RN15}, SimSiam \cite{RN16}, SwAV \cite{RN17}, and Barlow Twins \cite{RN18} were selected. The performance of this methodology shows variability depending on the types of data augmentation employed and the structure of the backbone network. Therefore, various experiments were conducted to identify the optimal approach for integrating EEG into contrastive learning-based SSL (detailed explanations are provided in Appendix C). For the masked prediction task-based SSL methodology, MAE \cite{RN20}, SimMIM \cite{RN22}, and Data2Vec \cite{RN23} were chosen. In contrast to the former methodologies, these methodologies do not involve the data augmentation process, and the backbone network remains fixed as ViT, thus making additional experimental procedures unnecessary. Raw EEG signals were transformed into short-time Fourier transform images and used as input data. For SSL for sleep staging, BENDR \cite{RN25}, ContraWR \cite{RN26}, TS-TCC \cite{RN28}, and mulEEG \cite{RN30} were selected, while for supervised learning for sleep staging, DeepSleepNet \cite{RN9}, IITNet \cite{RN10}, U-Sleep \cite{RN36}, AttnSleep \cite{RN11}, and SleepExpertNet \cite{RN12} were chosen.

\subsection{Experimental Setting}

\subsubsection{Evaluation Scheme}

The performance of the model was assessed via subject group k-fold cross-validation, as outlined \cite{RN28, RN30, RN27}. In detail, the construction of SSL-based methodologies involves dividing the dataset into three distinct groups: train, validation, and test datasets. The train dataset is utilized for SSL training and proceeds without labels. Subsequently, the validation dataset is utilized for linear evaluation and fine-tuning, leveraging a limited amount of labeled data. The test dataset is employed for comprehensive evaluation. In contrast, methodologies based on supervised learning are divided into train and test datasets, which are utilized for training and evaluation purposes. 

Additionally, this study employed three evaluation scenarios to compare the proposed model with another approach. For evaluation, it is necessary to attach a classifier network to the backbone network trained via SSL and then proceed with training using a few labeled data. This process is referred to as a downstream task. A detailed description of each evaluation protocol is as follows:

\begin{itemize}
  \item \textbf{(Evaluation Scenario 1, linear evaluation using single-epoch EEG)} The parameters of the backbone network are fixed, and then only the classifier network is trained. This method enables the evaluation of which SSL methodologies can effectively represent EEG features.
  \item \textbf{(Evaluation Scenario 2, fine-tuning using multi-epoch EEG)} The performance of the final model, which integrates the backbone network with TCM, referred to as NeuroNet+TCM, is evaluated. In this configuration, the backbone network is fixed, except for the last Transformer layer. This method not only facilitates additional training of the data's nonlinear features but also ensures enhanced performance due to the use of multi-epoch EEG. This evaluation scenario is utilized for comparison between the NeuroNet+TCM and supervised learning models.
  \item \textbf{(Evaluation Scenario 3, cross-dataset evaluation)} The performance of the proposed models (NeuroNet and NeuroNet+TCM) is evaluated using datasets that are different from the ones utilized for training. The aim is to determine whether these proposed models achieve outcomes comparable to or surpassing those achieved by supervised learning. For this, z-normalization was applied to the input signals to align the distributions of datasets. Additionally, models trained on each fold were combined using a soft-voting ensemble. This process was similarly applied to supervised learning models.
\end{itemize}

The impact of trainable parameter size on performance was examined using two models that share identical architecture yet have varying numbers of parameters. These models are designated as NeuroNet-B and NeuroNet-T, with NeuroNet-B possessing a greater number of model parameters than NeuroNet-T. Appendix A lists the hyperparameter values used in each evaluation scenario. These hyperparameter values were derived based on the results of ablation experiments.
    
\subsubsection{Evaluation Metric}
For overall performance measurement, overall accuracy (ACC) and macro-F1 score (MF1) were utilized, while per-class F1 score (F1) was used for measuring performance by class. Here, MF1 is a useful metric for evaluating model performance on imbalanced datasets. ACC and MF1 can be calculated if true positives (TP), false positives (FP), true negatives (TN), and false negatives (FN) for each class are provided. The formulas are as follows:
\begin{equation}
\label{equation_7}
ACC=\frac{\sum_{i=1}^{K}TP_{i}}{M}
\end{equation} 

\begin{equation}
\label{equation_8}
MF1=\frac{1}{K}\sum_{i=1}^{K}\frac{2\times Precision_{i}\times Recall_{i}}{Precision_{i}+Recall_{i}}
\end{equation}
Here, for each class, ${Precision}_i $ and ${Recall}_i $ are calculated as ${Precision}_i={{{TP}_i}/{({TP}_i+\;{FP}_i)}} $ and ${Recall}_i={{{TP}_i}/{({TP}_i+\;{FN}_i)}} $, respectively, where $M $ is the total number of samples, and $K $ is the number of classes. In this context, $K $ represents the 5 stages of sleep (Wake, N1, N2, N3, and REM).

\section{Results}

        \begin{table*}[!htbp]
        \caption{Comparison with other methodologies about linear evaluation using single-epoch EEG.}
        \label{table:table_2}
        \resizebox{\textwidth}{!}{
        \def\arraystretch{1}
        \ignorespaces 
        \centering \footnotesize 
        \renewcommand{\arraystretch}{1.15}
        \renewcommand{\tabcolsep}{1mm}
        \arrayrulecolor{black}
        \begin{tabular}{c|ccccc|cc|ccccc|cc|ccccc|cc} 
        \hline
                                                              & \multicolumn{7}{c|}{\textbf{\textit{Sleep-EDFX }}}                                                                          & \multicolumn{7}{c|}{\textbf{\textit{SHHS }}}                                                                                & \multicolumn{7}{c}{\textbf{\textit{ISRUC-Sleep }}}                                                                          \\ 
        \cline{2-22}
                                                              & \multicolumn{5}{c|}{\textit{Per-Class F1 }}                                        & \multicolumn{2}{c|}{\textit{Overall }} & \multicolumn{5}{c|}{\textit{Per-Class F1 }}                                        & \multicolumn{2}{c|}{\textit{Overall }} & \multicolumn{5}{c|}{\textit{Per-Class F1 }}                                        & \multicolumn{2}{c}{\textit{Overall }}  \\ 
        \cline{2-22}
                                                              & \textit{W}     & \textit{N1}    & \textit{N2}    & \textit{N3}    & \textit{REM}   & \textit{ACC}   & \textit{MF1}          & \textit{W}     & \textit{N1}    & \textit{N2}    & \textit{N3}    & \textit{REM}   & \textit{ACC}   & \textit{MF1}          & \textit{W}     & \textit{N1}    & \textit{N2}    & \textit{N3}    & \textit{REM}   & \textit{ACC}   & \textit{MF1}          \\ 
        \hline
        \textit{SimCLR}                                       & 85.95          & 32.25          & 79.44          & 68.65          & 53.59          & 71.49          & 63.98                 & 83.61          & 23.14          & 80.64          & 82.75          & 71.92          & 77.75          & 68.41                 & 82.24          & 39.41          & 69.38          & 78.43          & 68.31          & 70.98          & 67.55                 \\
        \textit{BYOL}                                         & 87.08          & 33.40          & 78.32          & 64.49          & 55.12          & 71.82          & 63.68                 & 83.20          & 19.50          & 81.71          & 84.52          & 71.32          & 78.51          & 68.05                 & 82.30          & 40.62          & 70.42          & 77.48          & 67.76          & 71.22          & 67.72                 \\
        \textit{SwAV}                                         & 85.11          & 32.49          & 78.92          & 62.68          & 54.44          & 71.98          & 62.73                 & 81.88          & 21.80          & 80.10          & 82.17          & 71.93          & 77.21          & 67.58                 & 80.22          & 38.50          & 68.00          & 75.60          & 66.15          & 68.91          & 65.69                 \\
        \textit{SimSiam}                                      & 85.55          & 29.30          & 78.92          & 63.62          & 46.89          & 71.34          & 60.85                 & 84.77          & 21.06          & 81.59          & 83.56          & 73.09          & 79.23          & 68.81                 & 81.55          & 38.05          & 68.74          & 76.13          & 66.05          & 69.68          & 66.10                 \\
        \textit{Barlow Twins}                                 & 87.85          & 29.20          & \textbf{81.84} & 69.69          & 57.33          & 75.13          & 65.18                 & 84.36          & 23.84          & 81.69          & 83.82          & 74.13          & 79.27          & 69.57                 & 83.84          & 40.02          & 71.24          & 78.65          & 68.68          & 72.23          & 68.49                 \\ 
        \hline
        \textit{MAE}                                          & 81.23          & 27.39          & 76.72          & 60.99          & 46.91          & 68.31          & 58.65                 & 85.28          & 18.38          & 83.95          & 84.90          & 75.86          & 81.08          & 69.67                 & 81.94          & 36.41          & 74.01          & 83.69          & 58.42          & 71.49          & 66.89                 \\
        \textit{SimMIM}                                       & 83.68          & 26.69          & 75.73          & 51.32          & 44.32          & 69.94          & 56.35                 & 84.94          & 22.38          & 83.14          & 85.11          & 75.25          & 80.45          & 70.16                 & 82.05          & 35.75          & 74.43          & 84.13          & 58.79          & 71.64          & 67.03                 \\
        \textit{Data2Vec}                                     & 83.39          & 28.54          & 75.41          & 56.54          & 50.82          & 69.78          & 58.94                 & 78.07          & 16.65          & 79.08          & 81.02          & 71.24          & 76.18          & 65.21                 & 82.22          & 36.59          & 73.42          & 83.09          & 60.83          & 71.71          & 67.23                 \\ 
        \hline
        \textit{BENDR}                                        & 72.15          & 28.28          & 67.83          & 50.94          & 32.50          & 57.42          & 50.34                 & 52.34          & 08.41          & 72.72          & 78.37          & 54.78          & 65.08          & 53.32                 & 52.75          & 13.83          & 72.18          & 78.28          & 55.69          & 64.78          & 54.55                 \\
        \textit{ContraWR}                                     & 88.40          & 34.35          & 81.67          & 68.81          & 62.78          & 75.79          & 67.20                 & 85.78          & 25.51          & 84.20          & 85.79          & 77.53          & 81.65          & 71.76                 & 84.09          & 40.23          & 73.26          & 82.55          & \textbf{71.18} & 74.07          & 70.26                 \\
        \textit{TS-TCC}                                       & 73.28          & 21.15          & 66.01          & 41.39          & 37.33          & 61.45          & 47.83                 & 70.05          & 17.68          & 75.33          & 73.23          & 62.00          & 70.43          & 59.66                 & 80.91          & 32.06          & 70.13          & 80.27          & 64.17          & 70.17          & 65.51                 \\
        \textit{mulEEG}                                       & 89.09          & \textbf{36.52} & 80.69          & 69.62          & 59.66          & 74.92          & 67.12                 & 83.67          & 21.41          & 83.06          & 85.82          & 74.16          & 79.94          & 69.62                 & 80.87          & 36.69          & 71.25          & 82.56          & 65.77          & 71.58          & 67.43                 \\ 
        \hline
        \rowcolor[rgb]{0.855,0.914,0.969} \textit{NeuroNet-T} & \textbf{89.90} & 30.21          & 81.51          & 71.39          & 59.51          & 76.26          & 66.50                 & 83.97          & 13.95          & 83.30          & 85.45          & 73.75          & 80.45          & 68.09                 & 84.51          & 39.40          & 76.15          & 84.68          & 67.63          & 75.12          & 70.47                 \\
        \rowcolor[rgb]{0.855,0.914,0.969} \textit{NeuroNet-B} & 89.17          & 36.24          & 81.74          & \textbf{69.97} & \textbf{63.82} & \textbf{76.74} & \textbf{68.19}        & \textbf{88.27} & \textbf{30.76} & \textbf{86.20} & \textbf{87.56} & \textbf{79.41} & \textbf{84.13} & \textbf{74.44}        & \textbf{85.08} & \textbf{42.11} & \textbf{76.84} & \textbf{85.74} & 71.04          & \textbf{76.47} & \textbf{72.16}        \\
        \hline
        \end{tabular}}
        \end{table*}

        \begin{table*}[!htbp]
        \caption{Comparision between supervised learning-based methodologies and NeuroNet+TCM.}
        \label{table:table_3}
        \resizebox{\textwidth}{!}{
        \def\arraystretch{1}
        \ignorespaces 
        \centering 
        \footnotesize 
        \renewcommand{\arraystretch}{1.15}
        \renewcommand{\tabcolsep}{1mm}
        \begin{tabular}{c|c|ccccc|cc|ccccc|cc|ccccc|cc} 
        \hline
        \multirow{3}{*}{}                                         & \multirow{3}{*}{\begin{tabular}[c]{@{}c@{}}\textbf{\textit{EEG} }\\\textbf{\textit{Epoch} }\end{tabular}} & \multicolumn{7}{c|}{\textit{\textbf{Sleep-EDFX}}}                                                                          & \multicolumn{7}{c|}{\textit{\textbf{SHHS}}}                                                                                & \multicolumn{7}{c}{\textit{\textbf{ISRUC-Sleep}}}                                                                          \\ 
        \cline{3-23}
                                                                  &                                                                                         & \multicolumn{5}{c|}{\textit{Per-Class F1}}                                         & \multicolumn{2}{c|}{\textit{Overall}} & \multicolumn{5}{c|}{\textit{Per-Class F1}}                                         & \multicolumn{2}{c|}{\textit{Overall}} & \multicolumn{5}{c|}{\textit{Per-Class F1}}                                         & \multicolumn{2}{c}{\textit{Overall}}  \\ 
        \cline{3-23}
                                                                  &                                                                                         & \textit{W}     & \textit{N1}    & \textit{N2}    & \textit{N3}    & \textit{REM}   & \textit{ACC}   & \textit{MF1}         & \textit{W}     & \textit{N1}    & \textit{N2}    & \textit{N3}    & \textit{REM}   & \textit{ACC}   & \textit{MF1}         & \textit{W}     & \textit{N1}    & \textit{N2}    & \textit{N3}    & \textit{REM}   & \textit{ACC}   & \textit{MF1}         \\ 
        \hline
        \textit{DeepSleepNet}                                     & 25                                                                                      & 90.84          & 35.56          & 81.42          & 68.59          & 68.02          & 77.49          & 68.89                & 83.84          & 18.87          & 83.55          & 84.67          & 76.80          & 81.02          & 69.55                & 81.55          & 38.25          & 68.90          & 81.17          & 62.17          & 69.84          & 66.41                \\
        \textit{IITNet}                                           & 10                                                                                      & 92.59          & 45.77          & 83.61          & 63.65          & 80.90          & 81.48          & 73.30                & 89.32          & 47.38          & 85.57          & 80.96          & 87.58          & 84.74          & 78.16                & 84.60          & 40.51          & 78.39          & 85.27          & 79.15          & 77.89          & 73.59                \\
        \textit{U-Sleep}                                          & 35                                                                                      & 92.71          & 47.72          & 84.65          & 65.23          & 82.44          & 82.42          & 74.55                & 89.56          & 48.08          & 86.53          & 81.76          & \textbf{88.82} & 85.59          & 78.95                & \textbf{86.34} & 44.16          & \textbf{79.07} & 85.38          & \textbf{81.48} & \textbf{78.89} & \textbf{75.29}       \\
        \textit{AttnSleep}                                        & 1                                                                                       & 91.49          & 40.44          & 83.84          & 72.28          & 72.18          & 79.68          & 72.05                & 87.28          & 28.13          & 83.83          & 85.39          & 77.72          & 81.98          & 72.47                & 84.54          & 42.41          & 75.81          & 83.45          & 69.92          & 75.72          & 71.23                \\
        \textit{SleepExpertNet}                                   & 20                                                                                      & 92.80          & 52.75          & 85.92          & 73.40          & 80.93          & 83.13          & 77.16                & \textbf{90.84} & 40.71          & 86.60          & 84.04          & 87.94          & 85.94          & 78.03                & 72.59          & 14.25          & 66.61          & 72.69          & 58.05          & 64.89          & 56.84                \\
        \rowcolor[rgb]{0.855,0.914,0.969} \textit{NeuroNet-T+TCM} & 20                                                                                      & 92.27          & 53.01          & 85.19          & 75.23          & 77.13          & 82.67          & 76.57                & 86.44          & 50.17          & 85.97          & 85.97          & 83.99          & 84.62          & 78.51                & 83.75          & 44.73          & 77.50          & 86.61          & 72.04          & 76.79          & 72.93                \\
        \rowcolor[rgb]{0.855,0.914,0.969} \textit{NeuroNet-B+TCM} & 20                                                                                      & \textbf{93.15} & \textbf{58.80} & \textbf{87.21} & \textbf{76.97} & \textbf{83.00} & \textbf{85.24} & \textbf{79.82}       & 89.05          & \textbf{55.29} & \textbf{88.09} & \textbf{86.48} & 87.25          & \textbf{86.88} & \textbf{81.23}       & 84.50          & \textbf{46.09} & 77.25          & \textbf{86.86} & 72.57          & 77.05          & 73.45                \\
        \hline
        \end{tabular}}
        \end{table*}
        
        \begin{table*}[!htbp]
        \caption{Cross-dataset evaluation experiment applied to different PSG datasets.}
        \label{table:table_4}
        \resizebox{\textwidth}{!}{
        \renewcommand{\arraystretch}{1.1}
        \begin{tabular}{c|c|cc|cc|cc|cc|cc|cc|cc} 
        \hline
        \multirow{2}{*}{}                                         & \multirow{2}{*}{\begin{tabular}[c]{@{}c@{}}\textbf{\textit{EEG}}\\\textbf{\textit{Epoch }}\end{tabular}} & \multicolumn{2}{c|}{\textbf{\textit{A} → \textit{B}}} & \multicolumn{2}{c|}{\textbf{\textbf{\textit{A}~→~\textit{C}}}} & \multicolumn{2}{c|}{\textbf{\textit{B} → \textit{A}}} & \multicolumn{2}{c|}{\textbf{\textit{B} → \textit{C}}} & \multicolumn{2}{c|}{\textbf{\textit{C} →\textit{ A}}} & \multicolumn{2}{c|}{\textbf{\textit{C} → \textit{B}}} & \multicolumn{2}{c}{\textbf{\textit{Average}}}  \\ 
        \cline{3-16}
                                                                  &                                                                                                          & \textit{ACC}   & \textit{MF1}                         & \textit{ACC}   & \textit{MF1}                                  & \textit{ACC}   & \textit{MF1}                         & \textit{ACC}   & \textit{MF1}                         & \textit{ACC}   & \textit{MF1}                         & \textit{ACC}   & \textit{MF1}                         & \textit{ACC}   & \textit{MF1}                  \\ 
        \hline
        \textit{U-Sleep}                                          & 35                                                                                                       & 56.50          & 46.22                                & 56.71          & 49.73                                         & 71.39          & 59.09                                & 70.08          & 64.72                                & \textbf{59.38} & 45.62                                & 60.08          & 54.36                                & 62.36          & 45.00                         \\
        \textit{SleepExpertNet}                                   & 20                                                                                                       & 58.63          & 48.65                                & 58.46          & 51.93                                         & 70.75          & 59.21                                & 71.19          & 65.75                                & 58.15          & \textbf{47.22}                       & 67.41          & 60.77                                & 64.10          & 55.59                         \\
        \rowcolor[rgb]{0.855,0.914,0.969} \textit{NeuroNet-T}                                       & 1                                                                                                        & 66.94          & 52.83                                & 61.17          & 52.71                                         & 79.44          & 69.57                                & 73.50          & 63.28                                & 50.18          & 42.21                                & 74.85          & 64.81                                & 67.68          & 57.56                         \\
        \rowcolor[rgb]{0.855,0.914,0.969} \textit{NeuroNet-B}                                       & 1                                                                                                        & 66.68          & 52.83                                & 61.06          & 53.19                                         & 79.80          & 70.40                                & 74.21          & 63.98                                & 50.20          & 42.14                                & 75.83          & 65.57                                & 67.93          & 58.01                         \\
        \rowcolor[rgb]{0.855,0.914,0.969} \textit{NeuroNet-T+TCM} & 20                                                                                                       & 73.82          & 63.20                                & \textbf{66.93} & \textbf{57.92}                                & 84.93          & 78.32                                & 82.09          & 75.29                                & 51.74          & 44.18                                & \textbf{83.01} & \textbf{75.66}                       & \textbf{73.75} & 65.76                         \\
        \rowcolor[rgb]{0.855,0.914,0.969} \textit{NeuroNet-B+TCM} & 20                                                                                                       & \textbf{73.86} & \textbf{64.12}                       & 65.29          & 56.32                                         & \textbf{85.12} & \textbf{79.98}                       & \textbf{83.95} & \textbf{76.59}                       & 50.93          & 42.03                                & 82.86          & 75.59                                & 73.67          & \textbf{65.77}                \\
        \hline
       \multicolumn{4}{l}{*\footnotesize{A: Sleep-EDFX, B: SHHS, C: ISRUC-Sleep}}
        \end{tabular}}        
        \end{table*}

\subsection{Comparison with State-of-the-Art Methodologies}

\subsubsection{Evaluation Scenario 1: Linear Evaluation using Single-epoch EEG}

NeuroNet-B demonstrated superior performance across three PSG datasets compared to other SSL methodologies (Table \ref{table:table_2}) in learning representations of EEG signals. Particularly for the SHHS, both class-specific and overall performances were outstanding. For the ISRUC-Sleep, it exhibited exceptional performance across all metrics except for the REM class. Even though the performance of NeuroNet-T exhibited a slight decrease in comparison to NeuroNet-B, it nonetheless showcased a noteworthy degree of efficacy. In the Sleep-EDFX, NeuroNet-T's performance in the Wake class was notably superior, and its overall accuracy was the highest, excluding NeuroNet-B. For the ISRUC-Sleep, NeuroNet-T exhibited higher overall performance compared to other SSL methodologies.

\subsubsection{Evaluation Scenario 2: Fine-tuning using Multi-epoch EEG}
On closer examination, when compared to the latest supervised learning methodologies, NeuroNet-B+TCM shows the highest performance across all metrics for the Sleep-EDFX (Table \ref{table:table_3}). In the SHHS, it also displays the highest performance in all classes except for the Wake and REM. The notable encouragement arises from the fact that supervised learning models harness an extensive array of labeled datasets. For the ISRUC-Sleep, it was observed that NeuroNet-B+TCM falls short in overall performance compared to U-Sleep \cite{RN36}. However, it still exhibits the highest performance in the N1 and N3 classes compared to other methodologies. Similar to evaluation scenario 1, NeuroNet-T shows lower performance than NeuroNet-B. Nonetheless, NeuroNet-T+TCM showcases performance on par with that of supervised learning methodologies. Specifically, in the N1 class, it shows the highest performance excluding NeuroNet-B+TCM. Additionally, in the SHHS, it is noted that NeuroNet-T+TCM, excluding NeuroNet-B+TCM, achieves the highest MF1.

\subsubsection{Evaluation Scenario 3: Cross-Dataset Evaluation}
Table \ref{table:table_4} presents the results of comparing the proposed models with the two most outstanding models among supervised learning methodologies (i.e., U-Sleep \cite{RN36}, SleepExpertNet \cite{RN12}). Upon closer inspection, it is observed that the proposed models demonstrate superior performance compared to existing supervised learning models. Despite utilizing only single-epoch EEG, NeuroNet exhibits better performance than supervised learning, and NeuroNet+TCM achieves remarkable performance by employing multiple EEG epochs and TCM. Notably, the results for models trained on SHHS (i.e., B\ensuremath{\rightarrow} A, B \ensuremath{\rightarrow} C) are outstanding, attributed to the larger dataset size of SHHS compared to other datasets. However, the performance of models trained on ISRUC-Sleep and evaluated on Sleep-EDFX (i.e., C \ensuremath{\rightarrow} A) is superior in supervised learning methodologies. Nonetheless, excluding C \ensuremath{\rightarrow} A, the proposed models outperform in all other aspects. It was shown that there are no noticeable performance gaps attributable to differences in model size. In conclusion, it has been confirmed that the proposed models showcase superior generalization performance when compared with supervised learning methodologies, effectively operating on datasets extending beyond the scope of their training dataset.

\subsection{Ablation Experiments}
An ablation experiment was conducted to derive the optimal settings information for the proposed model. Across all experiments, NeuroNet-B served as the base backbone, with Sleep-EDFX serving as the reference dataset. In light of the results obtained from these experiments, hyperparameters were established.

\subsubsection{Evaluation Scenario 1: Linear Evaluation using Single-epoch EEG}

\,\quad \textbf{(Frame Design)} The performance is evaluated based on various frame designs. Looking at Table \ref{table:table_5}, it indicates a trend where decreasing the frame size and overlap step generally leads to performance enhancements, but they significantly increase the training speed. Specifically, the configuration with a frame size of 3 and an overlap step of 0.375 achieves the highest performance, exhibiting an ACC of 76.74\% and a MF1 of 68.19\%. Therefore, in this study, the frame size and overlap step were fixed at 3 and 0.375, respectively. Considering the impact on training speed, the process of reducing these two values was omitted.

    \begin{table}[!htbp]
    \caption{Linear evaluation under different frame size and overlap size.}
    \label{table:table_5}
    \def\arraystretch{1}
    \ignorespaces 
    \centering 
    \begin{tabular}{c|c|c|c|c} 
    \hline
    \multicolumn{2}{c|}{\textbf{Frame Setting (sec) }}  & \multicolumn{2}{c|}{\textbf{Performance }} & \multirow{2}{*}{\begin{tabular}[c]{@{}c@{}}\textbf{Training Time}\\\textbf{/ Epoch (min)}\end{tabular}}  \\ 
    \cline{1-4}
    \textit{Frame Size}         & \textit{Overlap Step} & \textit{ACC}   & \textit{MF1}              &                                                                                                          \\ 
    \hline
    \multirow{3}{*}{\textbf{3}} & \textbf{0.375}        & \textbf{76.74} & \textbf{68.19}            & \textbf{20:33}                                                                                           \\
                                & 0.75                  & 76.52          & 67.98                     & 10:35                                                                                                    \\
                                & 1.5                   & 76.49          & 67.62                     & 05:46                                                                                                    \\ 
    \hline
    \multirow{3}{*}{4}          & 0.5                   & 76.36          & 67.26                     & 15:36                                                                                                    \\
                                & 1                     & 76.32          & 67.33                     & 07:47                                                                                                    \\
                                & 2                     & 76.42          & 67.14                     & 04:19                                                                                                    \\ 
    \hline
    \multirow{3}{*}{5}          & 0.625                 & 76.28          & 67.07                     & 11:34                                                                                                    \\
                                & 1.25                  & 76.65          & 67.38                     & 06:13                                                                                                    \\
                                & 2.5                   & 76.07          & 67.00                     & 03:35                                                                                                    \\ 
    \hline
    \multirow{3}{*}{6}          & 0.75                  & 76.12          & 67.01                     & 09:43                                                                                                    \\
                                & 1.5                   & 76.11          & 66.59                     & 05:08                                                                                                    \\
                                & 3                     & 75.95          & 66.46                     & 02:59                                                                                                    \\
    \hline
    \end{tabular}
    \end{table}

    \begin{figure}[!htbp]
        \centerline{\includegraphics[width=\columnwidth]{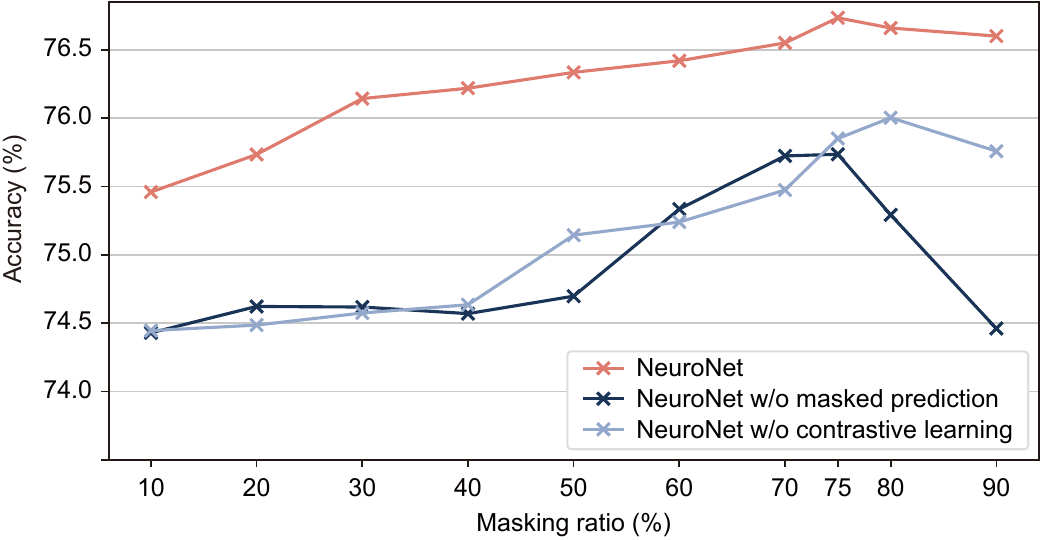}}
        \caption{Impact of different masking ratios on NeuroNet performance.}
        \label{fig:figure_3} 
        \vspace{0.4cm}
    \end{figure}

\textbf{(Masking Ratio)} At high masking ratio, NeuroNet demonstrates excellent performance. Figure \ref{fig:figure_3} illustrates the configurations for analyzing the effects on two tasks. NeuroNet w/o masked prediction shows highest accuracy at masking ratios of 70\% to 75\%, while NeuroNet w/o contrastive learning exhibits exceptional performance at ratios of 75\% to 90\%. NeuroNet, applying both tasks, outperforms single-task across all masking ratios, indicating their mutual complementarity. NeuroNet achieves its highest performance at a masking ratio of 75\%.

\textbf{(Decoder Depth and Width) }Upon reviewing the linear evaluation under different decoder dimensions and decoder depths (Table \ref{table:table_6}), NeuroNet tends to achieve superior performance when using smaller decoders. This is attributed to the characteristic of a smaller decoder, which requires a greater amount of semantic information to successfully accomplish the reconstruction task. Consequently, this necessitates the encoder to generate representations imbued with richer information, thereby enhancing the overall performance of the model. However, it was observed that if the decoder is too small, performing the reconstruction task becomes excessively challenging, leading to degraded performance. Therefore, NeuroNet achieves its highest performance when the dimension and depth of the decoder are set to 256 and 3, respectively.

\subsubsection{Evaluation Scenario 2: Fine-tuning using Multi-epoch EEG}
\,\quad \textbf{(Temporal Context Module)} The optimal structure for effectively analyzing temporal variations or correlations among multiple EEG epochs was explored (Table \ref{table:table_7}). It was observed that the Mamba-based structure outperforms the widely used LSTM or multi-head attention-based structures in previous studies. Furthermore, examining the performance of Mamba based on the context length revealed that the best performance is achieved when the context length is 20.

    \begin{table}[!htbp]
    \caption{Linear evaluation under different decoder dimensions and decoder depths.}
    \label{table:table_6}
    \def\arraystretch{1}
    \ignorespaces 
    \centering 
    \begin{tabular}{c|c|c|c|c} 
    \hline
    \multicolumn{2}{c|}{\textit{\textbf{Decoder}}} & \multicolumn{2}{c|}{\textit{\textbf{Performance}}} & \multirow{2}{*}{\begin{tabular}[c]{@{}c@{}}\textit{\textbf{Model Size}} \\\textit{\textbf{(MB)}}\end{tabular}}  \\ 
    \cline{1-4}
    \textit{Dim}                  & \textit{Depth} & \textit{ACC}   & \textit{MF1}                      &                                                                                                                 \\ 
    \hline
    \multirow{4}{*}{192}          & 1              & 76.05          & 66.83                             & 123.96                                                                                                          \\
                                  & 2              & 76.21          & 67.34                             & 125.74                                                                                                          \\
                                  & 3              & 76.23          & 67.24                             & 127.52                                                                                                          \\
                                  & 4              & 76.32          & 67.35                             & 129.30                                                                                                          \\ 
    \hline
    \multirow{4}{*}{\textbf{256}} & 1              & 76.05          & 66.83                             & 125.62                                                                                                          \\
                                  & 2              & 76.51          & 67.40                             & 128.78                                                                                                          \\
                                  & \textbf{3}     & \textbf{76.74} & \textbf{68.19}                    & \textbf{131.94}                                                                                                 \\
                                  & 4              & 75.94          & 66.56                             & 135.10                                                                                                          \\ 
    \hline
    \multirow{4}{*}{512}          & 1              & 76.03          & 66.65                             & 136.20                                                                                                          \\
                                  & 2              & 76.02          & 67.17                             & 148.81                                                                                                          \\
                                  & 3              & 76.02          & 66.48                             & 161.42                                                                                                          \\
                                  & 4              & 75.84          & 66.52                             & 174.03                                                                                                          \\
    \hline
    \end{tabular}
    \end{table}

    \begin{table}[!htbp]
    \caption{Comparison of modules and context lengths comprising temporal context module.}
    \label{table:table_7}
    \def\arraystretch{1}
    \ignorespaces 
    \centering 
    \begin{tabular}{c|c|c|c} 
    \hline
    \multirow{2}{*}{\textbf{\textit{Model }}}               & \multirow{2}{*}{\begin{tabular}[c]{@{}c@{}}\textbf{\textit{Context}}\\\textbf{\textit{Length}}\end{tabular}} & \multicolumn{2}{c}{\textbf{\textit{Performance}}}  \\ 
    \cline{3-4}
                                                            &                                                                                                              & \textit{ACC}   & \textit{MF1}                      \\ 
    \hline
    LSTM                                                    & 20                                                                                                           & 80.65          & 74.67                             \\
    Multi-Head
      Attention                                  & 20                                                                                                           & 81.56          & 75.27                             \\
    LSTM
      + Multi-Head Attention                           & 20                                                                                                           & 80.90          & 74.62                             \\
    \rowcolor[rgb]{0.855,0.914,0.969} Mamba-based TCM (ours)          & 10                                                                                                           & 83.74          & 77.75                             \\
    \rowcolor[rgb]{0.855,0.914,0.969} \textbf{Mamba-based TCM (ours)} & \textbf{20}                                                                                                  & \textbf{85.24} & \textbf{79.82}                    \\
    \rowcolor[rgb]{0.855,0.914,0.969} Mamba-based TCM (ours)          & 30                                                                                                           & 85.13          & 79.80                             \\
    \hline
    \end{tabular}
    \end{table}

\subsection{Hypnograms} Figure \ref{fig:figure_4} depicts the predicted results (i.e., hypnograms) for one subject from each of the three PSG datasets. The top row represents the labels annotated by sleep experts, the middle row corresponds to NeuroNet+TCM, and the bottom row to NeuroNet alone. The difference between the 2 figures is that the former shows results for NeuroNet-B and the latter for NeuroNet-T. In detail, it can be observed that across all three PSG datasets, the application of TCM yields results more closely aligned with those annotated by sleep experts. This underscores the significance of effectively incorporating temporal context information, thereby contributing to performance improvement. Moreover, it is clear that NeuroNet-B, despite its increased number of parameters, generally tends to show higher accuracy compared to NeuroNet-T.

    \begin{figure*}[!htbp]
        \centerline{\includegraphics[width=0.95\textwidth]{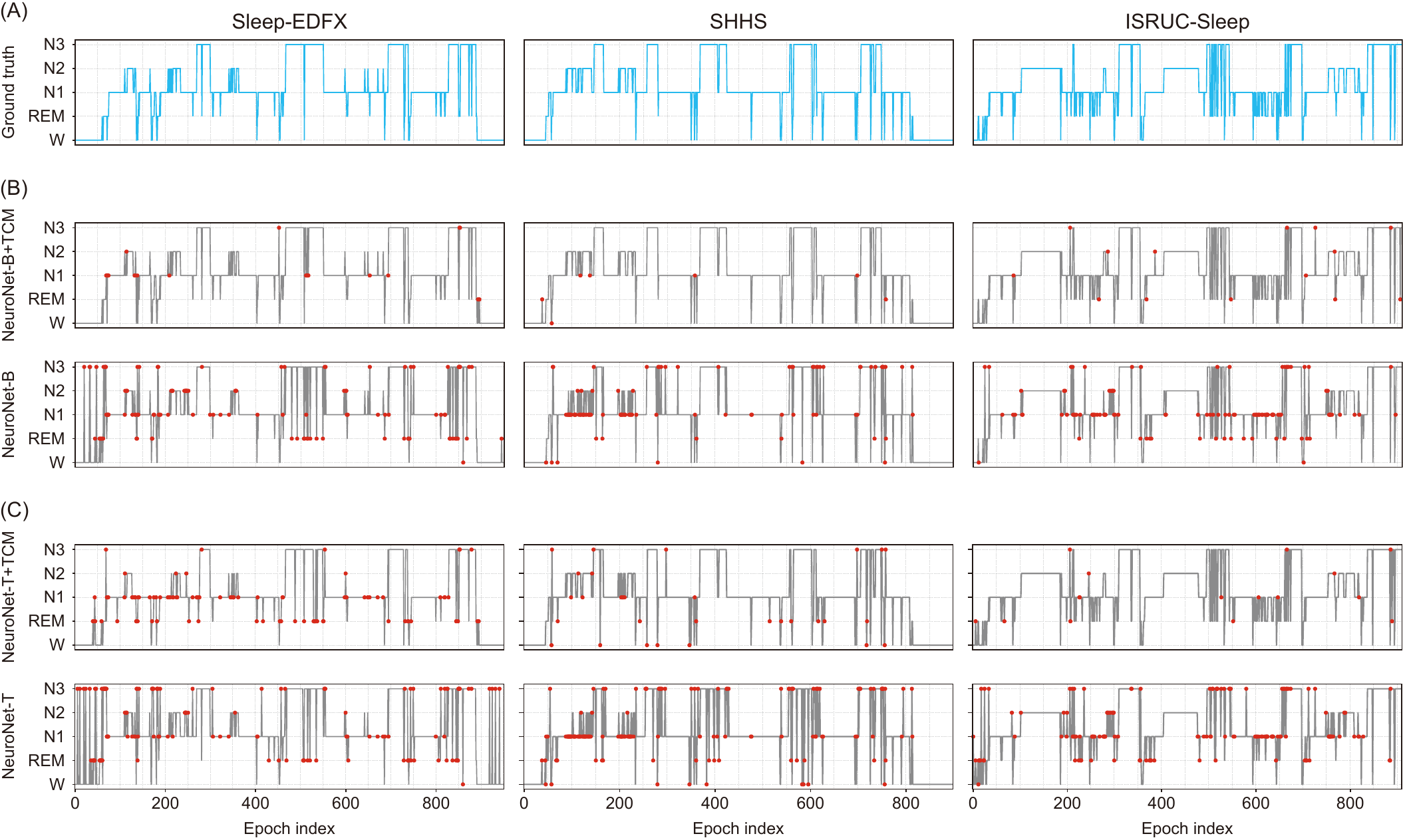}}
        \caption{The output hypnograms across five sleep stages. The first, second, and third columns correspond to \#sc4031e0, \#shhs1-204928, and \#subject-53 within Sleep-EDFX, SHHS, and ISRUC, respectively. (A) is manually scored by a sleep expert. (B) and (C) respectively represent NeuroNet-B and NeuroNet-T. The first row for both (B) and (C) displays the results for NeuroNet+TCM, while the second row shows the results for NeuroNet.~The errors are marked by the red\hspace{0.167em}dots.}
        \label{fig:figure_4} 
    \end{figure*}

\section{Discussions}
NeuroNet is a novel SSL framework that effectively combines the contrastive learning task with the masked prediction task. This study demonstrates that greater accuracy is achieved in NeuroNet performance when the contrastive learning task and masked prediction task are combined. It suggests that these two tasks mutually complement each other and lead to improved stability and the acquisition of higher-level representations. Consequently, NeuroNet showcases superior performance in comparison to the recent SSL methodologies (Table \ref{table:table_2}). Furthermore, it has been observed that when subjected to fine-tuning with a sparse number of labeled data, the NeuroNet+TCM configuration not only contends with but surpasses the performance of the latest supervised learning methodologies trained on substantially larger labeled data (Table \ref{table:table_3}, \ref{table:table_4}).

NeuroNet demonstrates exceptional performance and offers the advantage of omitting contrived EEG data augmentation for contrastive learning tasks. Generally, SSL methodologies based on contrastive learning necessitate the process of contrived data augmentation, posing a range of inherent challenges. Firstly, unlike images, the application of data augmentation to EEG signals risks diluting their intrinsic meaning. Secondly, selecting appropriate EEG data augmentation could be challenging, with suboptimal choices of data augmentation leading to ineffective SSL outcomes \cite{RN14, RN28, RN15}. Lastly, even when data augmentation is optimized for a specific model or dataset, there is a possibility it may not perform effectively with different models or datasets \cite{RN26, RN28, RN30}. NeuroNet inputs two subsets, randomly sampled differently, into the encoder, resulting in output vectors corresponding to different views, and then conducts a contrastive learning task on these vectors. This approach can be viewed as a challenge of determining whether the partially obscured portions of the entire EEG signal are the same or different. Compared to conventional methods, this approach is much simpler and effectively resolves the issues associated with contrived EEG data augmentation.

NeuroNet-B demonstrates higher performance compared to NeuroNet-T in most cases, primarily due to the relatively large scale of PSG datasets (Table \ref{table:table_3}, \ref{table:table_4}). The higher number of parameters enables the capture of more intricate and diverse patterns, thereby conferring an advantage in augmenting performance. Despite this, NeuroNet-T demonstrates superior performance over other SSL methodologies on the Sleep-EDFX and ISRUC-Sleep datasets and is competitive on the SHHS dataset. This suggests that although NeuroNet-T has fewer parameters, it still effectively captures the characteristics of EEG signals. Given the inherent challenges associated with EEG data acquisition, which often result in limited dataset capacity, leveraging NeuroNet-T with its fewer parameters may be a pragmatic strategy for optimal performance under such constraints.

The architecture of TCM, intricately designed to discern the relationships between different EEG epochs and based on the Mamba, emerged as a key driver for performance improvement. The overall performance exhibited a notable increase of approximately 4-5\% upon Mamba-based TCM, compared to the methodologies predominantly employed in prior studies, such as LSTM or multi-head attention-based methodologies (Table \ref{table:table_7}). Consequently, the model combining NeuroNet and the Mamba-based TCM (= NeuroNet+TCM), despite being trained on a limited amount of labeled data, demonstrated superior or comparable performance to supervised learning-based sleep staging trained on a vast amount of labeled data. This illustrates that the combination of NeuroNet, which effectively represents EEG features, and Mamba, specialized in intricate sequence modeling, yields highly efficacious outcomes. In particular, Mamba has addressed inefficiencies in long sequences and also improved performance by allowing the parameters of the SSM to be a function of the input.

Despite these advantages, there are still issues that need to be addressed. Firstly, SSL methodologies are designed to leverage unlabeled data by learning representations without explicit supervision. However, they do incorporate a small amount of labeled data at some point in the model development process. This means that the quality and accuracy of the initial labels may affect the performance of SSL-based methodologies. This can be particularly problematic during fine-tuning, where the quality of labels directly influences the model's ability to generalize from the learned representations to specific downstream tasks. This issue is especially pertinent in this study, which utilized public PSG data, where the unreliability of labels can and likely will be present. Therefore, future research will focus on conducting studies on "noisy label classification" optimized for sleep EEG signals to solve the issue of label reliability. Simultaneously, we plan to employ a large number of highly skilled sleep experts to select several support sets representing each sleep stage and then supplement them through few-shot or zero-shot learning to achieve accurate results without further training. Secondly, NeuroNet has been found to improve performance as the frame size and overlap size decrease, but this comes with a significant increase in computing cost (Table \ref{table:table_5}). This is because the core component of NeuroNet, the Transformer, struggles with efficiently processing samples with long sequences. Therefore, future research is expected to explore replacing Transformer with Mamba to achieve superior EEG representations along with more efficient computation, which would improve inference speed.

\appendices
\section{Training Settings and Hyperparameters}
The training and evaluation of the model were conducted on a computer equipped with an Intel I9-9980XE CPU at 3.00GHz, 128GB RAM, and an NVIDIA GPU 3090. Furthermore, all data processing and algorithm development was carried out using Python version 3.9, with the Pytorch version 1.10 library being utilized. The detailed hyperparameters are described in Table \ref{table:appendix_table_1}. Evaluation scenario 3 shares the same hyperparameters as scenario 2, as it does not involve an additional training process.

\begin{table}[!htbp]
\setcounter{table}{0}
\renewcommand{\thetable}{\Alph{section}\arabic{table}}
\centering
\caption{Hyperparameters for evaluation scenario for NeuroNet.}
\label{table:appendix_table_1}
\begin{tabular}{c|cc} 
\hline
\multirow{2}{*}{}       & \multicolumn{1}{c|}{\textbf{\textit{Scenario 1}}} & \textbf{\textit{Scenario 2}}  \\ 
\cline{2-3}
                        & \multicolumn{2}{c}{\textit{Self-Supervised Learning}}                             \\ 
\hline
epoch                   & \multicolumn{2}{c}{50}                                                            \\
batch size              & \multicolumn{2}{c}{1024}                                                          \\
frame size              & \multicolumn{2}{c}{3}                                                             \\
overlap step            & \multicolumn{2}{c}{0.75}                                                          \\
encoder dim             & \multicolumn{2}{c}{T: 512 / B: 768}                                               \\
encoder depth           & \multicolumn{2}{c}{T: 4 / B: 4}                                                   \\
encoder head            & \multicolumn{2}{c}{8}                                                             \\
decoder dim             & \multicolumn{2}{c}{T: 192 / B: 256}                                               \\
decoder depth           & \multicolumn{2}{c}{T: 1 / B: 3}                                                   \\
decoder head            & \multicolumn{2}{c}{8}                                                             \\
projection hidden       & \multicolumn{2}{c}{(1024, 512)}                                                   \\
temperature scale       & \multicolumn{2}{c}{0.5}                                                           \\
mask ratio              & \multicolumn{2}{c}{\textasciitilde{}}                                             \\
optimizer               & \multicolumn{2}{c}{AdamW}                                                         \\
optimizer momentum      & \multicolumn{2}{c}{(0.9, 0.999)~}                                                 \\
learning rate           & \multicolumn{2}{c}{2e-05}                                                         \\ 
\hline
                        & \multicolumn{2}{c}{\textit{Downstream Task}}                                      \\ 
\hline
epoch                   & 300                                               & 100                           \\
batch size              & 512                                               & 128                           \\
optimizer               & AdamW                                             & AdamW                         \\
optimizer momentum      & (0.9, 0.999)~                                     & (0.9, 0.999)~                 \\
learning rate           & 1e-05                                             & 5e-03                         \\
temporal context length & -                                                 & 20                            \\
mamba\_d\_state         & -                                                 & 16                            \\
mamba\_d\_conv          & -                                                 & 4                             \\
mamba\_expand           & -                                                 & 2                             \\
\hline
\end{tabular}
\end{table}

\section{Contrastive Learning based SSL with Single-Channel EEG}
SSL methodologies based on contrastive learning, such as SimCLR \cite{RN14}, BYOL \cite{RN15}, SwAV \cite{RN17}, SimSiam \cite{RN16}, Barlow Twins \cite{RN18}, etc., have demonstrated remarkable performance in the field of computer vision. However, compared to SSL methodologies based on masked prediction tasks (e.g., MAE \cite{RN20}, Data2Vec \cite{RN23}, etc.), their representation performance can vary significantly across different scenarios, such as data augmentation techniques and types of backbone networks. Thus, in this research, the following steps were taken to effectively apply contrastive learning-based SSL methodologies to single-channel EEG.

\subsection{Data Augmentation}
For effective training using contrastive learning-based SSL methodologies, selecting appropriate data augmentations is crucial \cite{RN14, RN37}. Inspired by \cite{RN38, RN37, RN28}, this research implemented five data augmentations optimized for single-channel EEG. During training, two data augmentations were randomly selected and applied to each data sample. Detailed descriptions of each data augmentation follow.

\begin{itemize}
  \item \textbf{(Random Gaussian Noise)} Adds Gaussian noise to the original signal.
  \item \textbf{(Random Crop)} Randomly crops a portion of the original signal and then interpolates it back to the original signal size.
  \item \textbf{(Random Bandpass Filtering)} Selects a frequency band at random and applies band-pass filtering to the original signal.
  \item \textbf{(Random Temporal Cutout)} Randomly selects a segment of the original signal and replaces it with the mean value of the original signal.
  \item \textbf{(Random Permutation)} Segments the signal randomly, shuffles these segments in a random order, and then merges them.
\end{itemize}

\subsection{Backbone Network}
To compare and analyze the performance of SSL methodologies based on different backbone networks, three backbone networks were selected and implemented. Each selected backbone network is a deep learning algorithm designed to analyze sleep stages using single-epoch EEG as input. Detailed descriptions are as follows:

\begin{itemize}
 \item \textbf{(DeepSleepNet \cite{RN9})} Utilizes two CNNs with different kernel sizes to extract low- and high-frequency features. Each CNN comprises four convolution layers and two max pooling layers. For this study, the bi-LSTM component of DeepSleepNet, typically used for processing single-epoch EEG, was omitted to solely focus on input from a single epoch.
 \item \textbf{(IITNet, intra-epoch version \cite{RN10})} Designed to process a single imbued EEG by dividing it into sub-epochs at fixed intervals, which are then fed into a ResNet to extract a representation vector. This vector is subsequently used as the input for a bi-LSTM, enabling the capture of temporal context.
 \item \textbf{(STFT Encoder \cite{RN26, RN30})} This model is designed to convert raw EEG signals into short-time Fourier transform spectrograms, which are then fed into four ResNet to facilitate the learning of sleep EEG signal features. This model has been widely adopted as a backbone model for SSL 
\end{itemize}

\subsection{Results}
Table \ref{table:appendix_table_2} presents the results of conducting 5-fold cross-validation on the Sleep-EDFX and SHHS, and 10-fold cross-validation on the ISRUC-Sleep dataset. Detailed examination reveals that the 'STFT Encoder' demonstrated the highest performance across all datasets, with the exception of Sleep-EDFX. Compared to other SSL methodologies, the Barlow Twins \cite{RN18} method showed superior performance overall.

\begin{table*}[!htbp]
\renewcommand{\thetable}{\Alph{section}\arabic{table}}
\caption{Linear evaluation based contrastive learning using single-epoch EEG.}
\label{table:appendix_table_2}
\resizebox{\textwidth}{!}{
\centering
\renewcommand{\arraystretch}{1.1}
\renewcommand{\tabcolsep}{0.7mm}
\arrayrulecolor{black}
\begin{tabular}{c!{\color{black}\vrule}c!{\color{black}\vrule}ccccc!{\color{black}\vrule}cc!{\color{black}\vrule}ccccc!{\color{black}\vrule}cc!{\color{black}\vrule}ccccc!{\color{black}\vrule}cc} 
\hline
\multicolumn{2}{c!{\color{black}\vrule}}{\multirow{3}{*}{\textit{~}}}                 & \multicolumn{7}{c!{\color{black}\vrule}}{\textbf{\textit{Sleep-EDFX}}}                                                                                             & \multicolumn{7}{c!{\color{black}\vrule}}{\textbf{\textit{SHHS}}}                                                                                & \multicolumn{7}{c}{\textbf{\textit{ISRUC-Sleep}}}                                                                          \\ 
\cline{3-23}
\multicolumn{2}{c!{\color{black}\vrule}}{}                                            & \multicolumn{5}{c!{\color{black}\vrule}}{\textit{Per-Class F1}\textbf{}}                              & \multicolumn{2}{c!{\color{black}\vrule}}{\textit{Overall}} & \multicolumn{5}{c!{\color{black}\vrule}}{\textit{Per-Class F1}}                    & \multicolumn{2}{c!{\color{black}\vrule}}{\textit{Overall}} & \multicolumn{5}{c!{\color{black}\vrule}}{\textit{Per-Class F1}}                    & \multicolumn{2}{c}{\textit{Overall}}  \\ 
\cline{3-9}\arrayrulecolor{black}\cline{10-13}\arrayrulecolor{black}\cline{14-16}\arrayrulecolor{black}\cline{17-20}\arrayrulecolor{black}\cline{21-21}\arrayrulecolor{black}\cline{22-23}
\multicolumn{2}{c!{\color{black}\vrule}}{}                                            & \textit{W}     & \textit{N1}    & \textit{N2}\textbf{} & \textit{N3}\textbf{} & \textit{REM}\textbf{} & \textit{ACC}\textbf{} & \textit{MF1}                       & \textit{W}     & \textit{N1}    & \textit{N2}    & \textit{N3}    & \textit{REM}   & \textit{ACC}   & \textit{MF1}                              & \textit{W}     & \textit{N1}    & \textit{N2}    & \textit{N3}    & \textit{REM}   & \textit{ACC}   & \textit{MF1}         \\ 
\arrayrulecolor{black}\hline
\multirow{3}{*}{\textit{SimCLR}}                                                & \textit{DeepSleepNet} & 84.70          & 26.41          & \textbf{81.61}       & \textbf{69.01}       & \textbf{54.54}        & \textbf{73.32}        & 63.25                              & 72.69          & 00.00          & 78.43          & 82.65          & 64.76          & 74.13          & 59.71                                     & 78.88          & 33.41          & \textbf{70.24} & \textbf{79.88} & 56.33          & 68.74          & 63.75                \\
                                                                       & \textit{IITNet}       & \textbf{85.95} & 32.25          & 79.44                & 68.65                & 53.59                 & 71.49                 & \textbf{63.98}                     & 81.63          & 14.07          & 79.09          & 79.85          & 67.27          & 75.53          & 64.38                                     & 74.17          & 26.93          & 63.95          & 76.59          & 53.33          & 62.87          & 58.99                \\
                                                                       & \textit{STFT Encoder} & 85.14          & \textbf{34.33} & 78.72                & 65.31                & 55.72                 & 71.97                 & 63.84                              & \textbf{83.61} & \textbf{23.14} & \textbf{80.64} & \textbf{82.75} & \textbf{71.92} & \textbf{77.75} & \textbf{68.41}                            & \textbf{82.24} & \textbf{39.41} & 69.38          & 78.43          & \textbf{68.31} & \textbf{70.98} & \textbf{67.55}       \\ 
\cline{1-17}\arrayrulecolor{black}\cline{18-20}\arrayrulecolor{black}\cline{21-21}\arrayrulecolor{black}\cline{22-23}
\multirow{3}{*}{\textit{BYOL}}                                                  & \textit{DeepSleepNet} & 86.41          & 31.14          & \textbf{81.16}       & \textbf{68.85}       & 45.78                 & \textbf{72.71}        & 62.67                              & 78.54          & 00.57          & 79.81          & 83.83          & 68.34          & 76.54          & 62.22                                     & 78.93          & 33.34          & \textbf{70.71} & 78.03          & 60.38          & 69.09          & 64.28                \\
                                                                       & \textit{IITNet}       & \textbf{87.08} & \textbf{33.40} & 78.32                & 64.49                & \textbf{55.12}        & 71.82                 & \textbf{63.68}                     & 83.20          & \textbf{19.50} & 81.71          & \textbf{84.52} & 71.32          & 78.51          & \textbf{68.05}                            & 79.44          & 31.35          & 68.31          & \textbf{78.23} & 62.45          & 68.13          & 63.96                \\
                                                                       & \textit{STFT Encoder} & 85.84          & 30.10          & 79.13                & 65.26                & 47.52                 & 71.69                 & 61.57                              & \textbf{85.06} & 07.16          & \textbf{82.00} & 83.93          & \textbf{73.36} & \textbf{79.91} & 66.30                                     & \textbf{82.30} & \textbf{40.62} & 70.42          & 77.48          & \textbf{67.76} & \textbf{71.22} & \textbf{67.72}       \\ 
\arrayrulecolor{black}\cline{1-21}\arrayrulecolor{black}\cline{22-23}
\multirow{3}{*}{\textit{SwAV}}                                                  & \textit{DeepSleepNet} & 83.62          & 24.07          & \textbf{81.54}       & \textbf{68.38}       & 53.90                 & \textbf{72.71}        & 62.30                              & 63.85          & 00.00          & 75.27          & 80.71          & 60.53          & 70.05          & 56.07                                     & 79.40          & 35.88          & \textbf{69.46} & \textbf{77.00} & 57.75          & 68.08          & 63.90                \\
                                                                       & \textit{IITNet}       & \textbf{85.45} & 32.25          & 79.09                & 65.91                & 50.40                 & 70.59                 & 62.62                              & 79.83          & 13.07          & 76.43          & 76.49          & 66.10          & 72.95          & 62.38                                     & 71.05          & 24.88          & 63.37          & 74.17          & 50.88          & 61.38          & 56.87                \\
                                                                       & \textit{STFT Encoder} & 85.11          & \textbf{32.49} & 78.92                & 62.68                & \textbf{54.44}        & 71.98                 & \textbf{62.73}                     & \textbf{81.88} & \textbf{21.80} & \textbf{80.10} & \textbf{82.17} & \textbf{71.93} & \textbf{77.21} & \textbf{67.58}                            & \textbf{80.22} & \textbf{38.50} & 68.00          & 75.60          & \textbf{66.15} & \textbf{68.91} & \textbf{65.69}       \\ 
\arrayrulecolor{black}\cline{1-22}\arrayrulecolor{black}\cline{23-23}
\multirow{3}{*}{\textit{SimSiam}}                                               & \textit{DeepSleepNet} & 81.45          & 23.97          & 76.46                & \textbf{66.93}       & 28.00                 & 67.79                 & 55.36                              & 71.44          & 00.06          & 73.71          & 73.84          & 57.87          & 69.20          & 55.38                                     & 78.36          & 34.81          & \textbf{71.20} & \textbf{79.17} & 61.16          & 69.47          & 64.94                \\
                                                                       & \textit{IITNet}       & 83.98          & \textbf{29.55} & 77.14                & 61.75                & 46.81                 & 69.05                 & 59.85                              & 72.48          & 01.75          & 74.10          & 79.47          & 59.11          & 70.31          & 57.38                                     & 75.23          & 24.97          & 63.35          & 73.06          & 56.51          & 63.20          & 58.62                \\
                                                                       & \textit{STFT Encoder} & \textbf{85.55} & 29.30          & \textbf{78.92}       & 63.62                & \textbf{46.89}        & \textbf{71.34}        & \textbf{60.85}                     & \textbf{84.77} & \textbf{21.06} & \textbf{81.59} & \textbf{83.56} & \textbf{73.09} & \textbf{79.23} & \textbf{68.81}                            & \textbf{81.55} & \textbf{38.05} & 68.74          & 76.13          & \textbf{66.05} & \textbf{69.68} & \textbf{66.10}       \\ 
\arrayrulecolor{black}\cline{1-22}\arrayrulecolor{black}\cline{23-23}
\multirow{3}{*}{\begin{tabular}[c]{@{}c@{}}\textit{Barlow}\\\textit{Twins}\end{tabular}} & \textit{DeepSleepNet} & \textbf{87.85} & 29.20          & \textbf{81.84}       & \textbf{69.69}       & \textbf{57.33}        & \textbf{75.13}        & \textbf{65.18}                     & 77.70          & 00.00          & 80.21          & \textbf{83.87} & 67.68          & 76.58          & 61.89                                     & 80.97          & 36.49          & \textbf{71.27} & \textbf{79.64} & 64.29          & 70.80          & 66.53                \\
                                                                       & \textit{IITNet}       & 86.46          & 28.99          & 77.99                & 61.75                & 52.97                 & 70.98                 & 61.63                              & 82.94          & 19.59          & 80.95          & 83.34          & 72.01          & 78.04          & 67.77                                     & 78.92          & 27.91          & 67.16          & 77.65          & 57.96          & 66.74          & 61.92                \\
                                                                       & \textit{STFT Encoder} & 86.03          & \textbf{31.59} & 80.40                & 66.12                & 55.48                 & 73.18                 & 63.92                              & \textbf{84.36} & \textbf{23.84} & \textbf{81.69} & 83.82          & \textbf{74.13} & \textbf{79.27} & \textbf{69.57}                            & \textbf{83.84} & \textbf{40.02} & 71.24          & 78.65          & \textbf{68.68} & \textbf{72.23} & \textbf{68.49}       \\
\arrayrulecolor{black}\hline
\end{tabular}}
\end{table*}

\section{Normalized Confusion Matrices}
\begin{figure}[!htbp]
        \centerline{\includegraphics[width=\columnwidth]{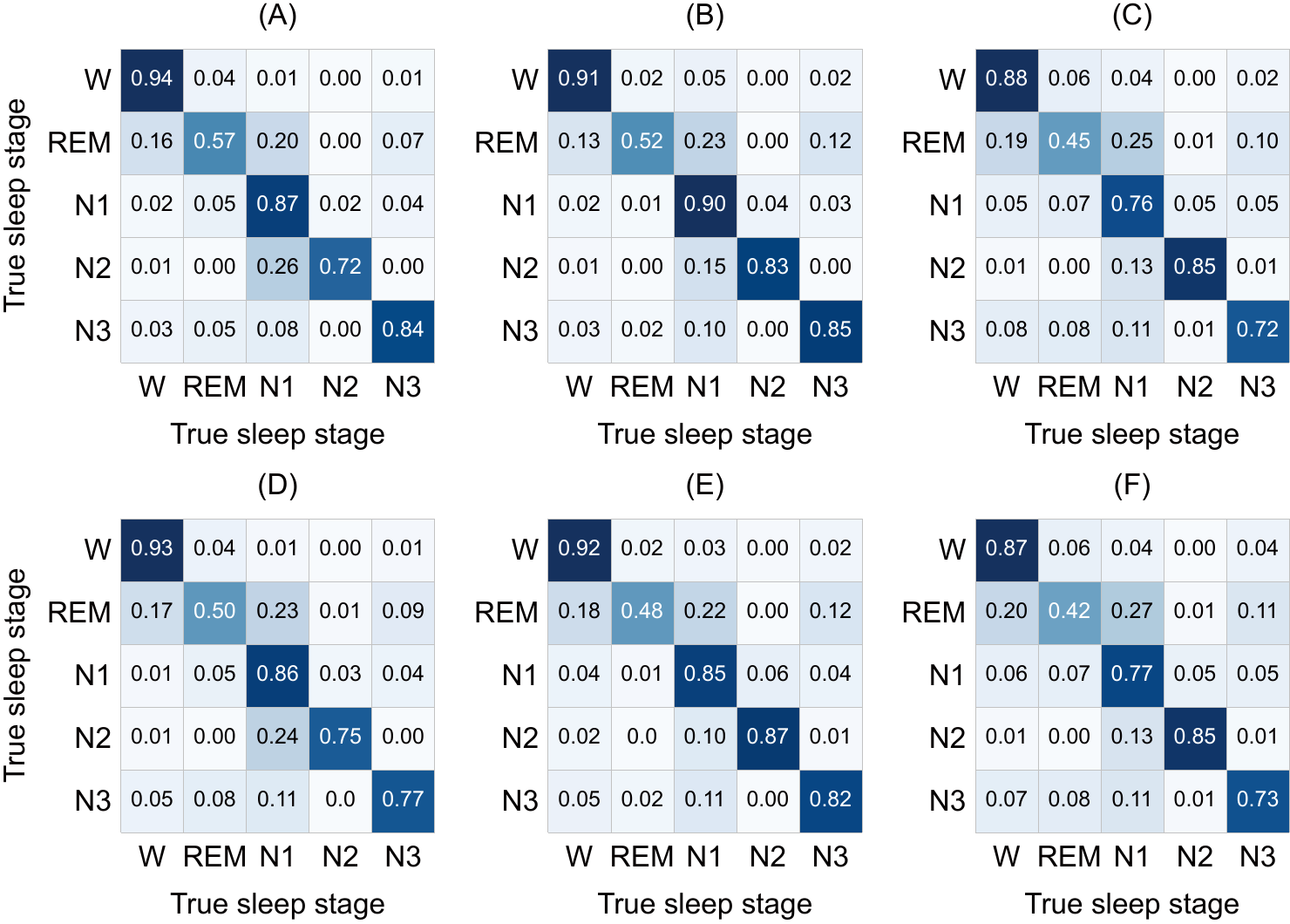}}
        \caption{The confusion matrices for sleep stage classification from evaluation scenario 2. The columns correspond to Sleep-EDFX, SHHS, and ISRUC-Sleep, respectively. Moreover, the first row signifies NeuroNet-B+TCM, and the second row depicts NeuroNet-T+TCM.}
        \label{fig:appendix_confusion_matrix_2} 
    \end{figure}

Appendix Figures \ref{fig:appendix_confusion_matrix_1} and \ref{fig:appendix_confusion_matrix_2} display the normalized confusion matrices. The distinction lies in that Figure \ref{fig:appendix_confusion_matrix_2} presents results with the TCM, unlike Figure \ref{fig:appendix_confusion_matrix_1}. Examination of both figures reveals that the application of TCM contributes to an overall enhancement in performance, with NeuroNet-B demonstrating greater accuracy compared to NeuroNet-T.

    \begin{figure}[!htbp]
    \setcounter{figure}{0}
        \centerline{\includegraphics[width=\columnwidth]{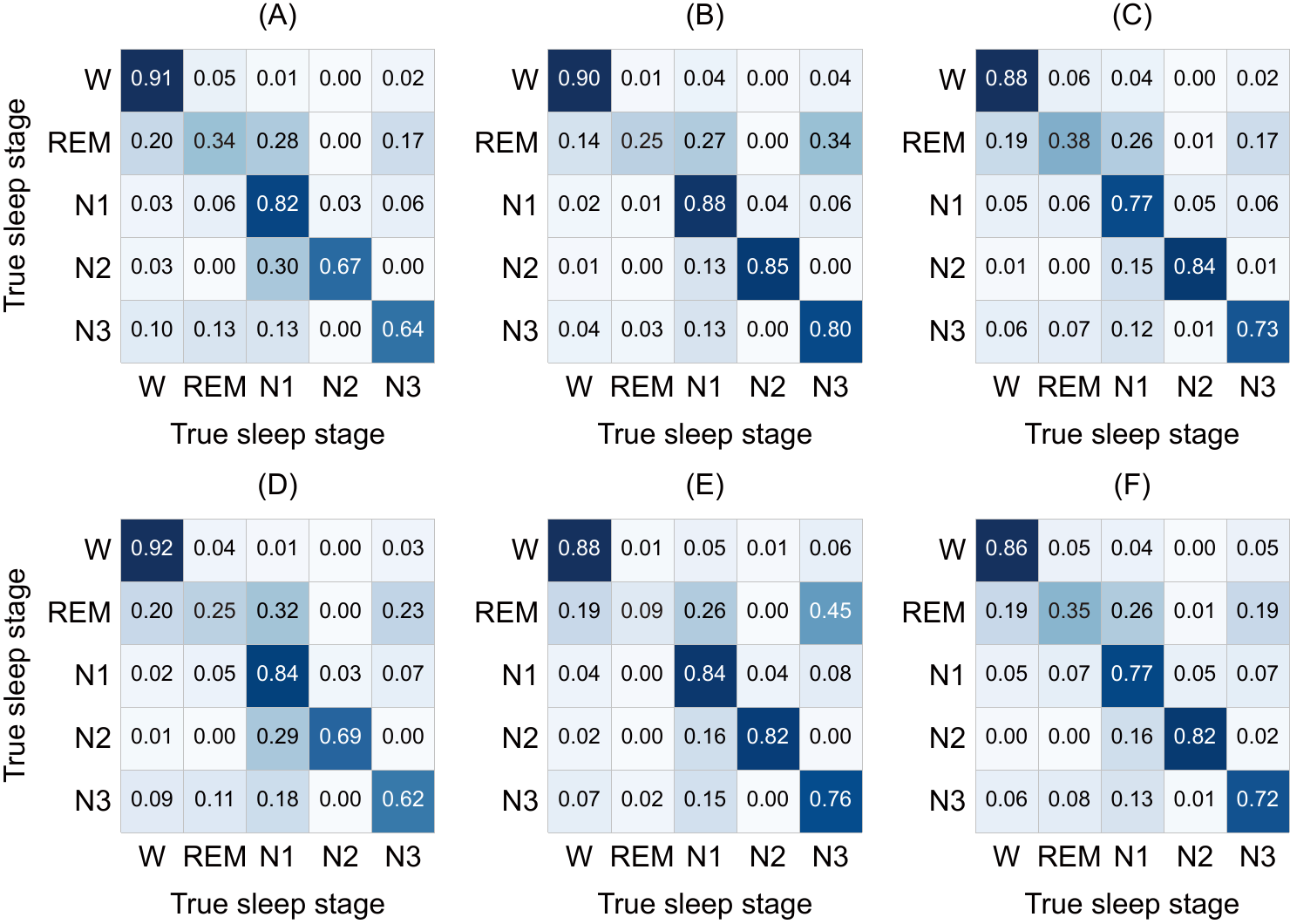}}
        \caption{The confusion matrices for sleep stage classification from evaluation scenario 1. The columns correspond to Sleep-EDFX, SHHS, and ISRUC-Sleep, respectively. Moreover, the first row signifies NeuroNet-B, and the second row depicts NeuroNet-T.}
        \label{fig:appendix_confusion_matrix_1} 
    \end{figure}
    
\ifCLASSOPTIONcompsoc
  \section*{Acknowledgments}
\else
  \section*{Acknowledgment}
\fi

We extend our sincere gratitude to Hyun-Ku Kang for his diverse contributions during the initial phases of our work; This work was supported by a National Research Foundation of Korea (NRF) Grant funded by the Korean government (Ministry of Science and ICT, MSIT) (No. 2022R1A2C1013205); 

\ifCLASSOPTIONcaptionsoff
  \newpage
\fi

\bibliographystyle{IEEEtran}

\bibliography{reference}

\end{document}